\documentclass[a4paper,11pt]{article}

\usepackage{jcappub} 

\usepackage[T1]{fontenc} 
\usepackage{subfig}
\usepackage{lscape}
\usepackage{cleveref}
\usepackage{graphicx}
\usepackage{xspace}
\usepackage{units}
\usepackage{xcolor}
\usepackage{subfig}
\usepackage{multirow}

\title{Upper limits on the total cosmic-ray luminosity of individual sources}

\author[a,1]{R.C. Anjos,\note{Corresponding author.}}
\author[a]{V. de Souza}
\author[b]{A.D. Supanitsky}

\affiliation[a]{Instituto de F\'{\i}sica de S\~ao Carlos, Universidade de S\~ao Paulo, Brazil}
\affiliation[b]{Instituto de Astronom\'ia y F\'{\i}sica del Espacio (IAFE), CONICET-UBA, Argentina}

\emailAdd{rita@ifsc.usp.br}
\emailAdd{vitor@ifsc.usp.br}
\emailAdd{supanitsky@iafe.uba.ar}

\abstract{In this paper, upper limits on the total luminosity of ultra-high-energy cosmic rays (UHECR)
($E > 10^{18}$ eV) are determined for five individual sources. The upper limit on the integral flux of GeV-TeV
gamma-rays of a given source is used to extract the upper limit on the total UHECR luminosity. The correlation
between upper limit on the integral GeV-TeV gamma-ray flux and upper limit on the UHECR luminosity is established
through the cascading process that takes place during propagation of the cosmic rays in the background radiation
fields, as explained in reference~\cite{bib:model}. Twenty-eight sources measured by FERMI-LAT, VERITAS and MAGIC
observatories have been studied. The measured upper limit on the GeV-TeV gamma-ray flux is restrictive enough to allow
the calculation of an upper limit on the total UHECR cosmic-ray luminosity of five sources. The upper limit on the
UHECR cosmic-ray luminosity of these sources is shown for several assumptions on the emission mechanism. For all studied
sources an upper limit on the ultra-high-energy proton luminosity is also set.}

\begin{document}
\maketitle
\flushbottom

\section{Introduction}
\label{sec:intro}

The observations of UHECR lack information on individual sources. The flux of particles with energy beyond $10^{18}$ eV
is extremely low and even large area experiments have not been able to measure enough events in order to identify or study
individual sources. Beyond that, charged particles deviate in the poorly known galactic and intergalactic magnetic fields
such that the arrival direction of the particle reaching Earth might not point back to its source. This caveat restricts the
study of UHECR individual sources.

In this work a method first introduced in reference~\cite{bib:model} is used to extract an upper limit on the UHECR luminosity
of individual sources from upper limits on the integral flux of GeV-TeV gamma-rays. The underlying idea is based on the measured
upper limit on the integral GeV-TeV gamma-ray flux to limit the UHECR luminosity. GeV-TeV gamma-rays are produced in the
propagation of UHECR in the intergalactic medium due to the interaction of the particles with the radiation backgrounds:
The Cosmic Microwave Background (CMB) and the Extragalactic Background Light (EBL).

In the original proposal, reference~\cite{bib:model}, the method was explained and two sources were studied in detail
(Pictor A and NGC 7469). For both sources, the measured upper limit on the integral flux of GeV-TeV gamma-ray was not strict
enough in order to constrain the total UHECR luminosity. However, the upper limits on the integral flux of GeV-TeV
gamma-rays were used to limit the proton luminosity. In this paper twenty eight sources are analyzed. For each of
them an upper limit on the proton luminosity is set. For five sources also an upper limit on the total UHECR luminosity
is set. Further studies with more sources are under way.

The calculated upper limits on the UHECR are of the order of the main proposed acceleration models ($10^{44} - 10^{46}$
erg s$^{-1}$)~\cite{bib:models}. The future Cherenkov Telescope Array (CTA)~\cite{bib:cta} Observatory will have
a sensitivity one order of magnitude smaller than the one corresponding to the current gamma-ray observatories. The upper limits
measured by CTA will allow a calculation of upper limits on the UHECR luminosity in the order of $10^{42}$ erg s$^{-1}$ for many
sources, depending on the energy threshold of the gamma-ray measurement. This will allow an unprecedented study of UHECR sources
by using this technique.

This paper is organized as follows. Section~\ref{sec:source:sel} explains how to select sources with relevant gamma-ray data in
order to set a valid upper limit on the UHECR luminosity. In section~\ref{sec:upper} the proton and total UHECR luminosity of
several sources are calculated according to the assumptions detailed in appendix~\ref{sec:iron:max}. Section~\ref{sec:conclusion}
summarizes the final remarks of this paper.

\section{Source selection}
\label{sec:source:sel}

The data measured by the FERMI-LAT~\cite{bib:fermi:cat:1,bib:fermi:cat:sey}, MAGIC~\cite{bib:magic} and VERITAS~\cite{bib:veritas}
telescopes have been searched. All gamma-ray sources with redshift smaller than 0.048 ($\sim$ 200 Mpc) have been selected. Among those,
twenty eight sources have been selected according to the procedure explained in this section. These sources are listed in~\cref{tab:1,tab:2,tab:3}.
All sources are active galactic nuclei (AGN), three are classified as radio galaxies and twenty one as Seyfert galaxies. The maximum distant
has been chosen in order to limit the data set to be analyzed within the horizon of most probable distance for an UHECR source. Further studies
with more distant sources are under way.

The observed cosmic-ray flux imposes an obvious upper limit on the cosmic-ray luminosity of any individual source, i.e. the contribution
of a given source cannot exceed the cosmic-ray flux of all sources observed by an UHECR experiment. So, for a given emission scenario, as
a function of distance of the source, how low must the upper limit on the integral flux of GeV-TeV gamma-ray flux of an individual source
($I^{UHECR}_{\gamma}$) be in order not to exceed the UHECR flux measured by a given experiment?

In order to answer this question let us consider the following argument. For a source at a given comoving distance ($D_s$) from Earth,
assuming an isotropic cosmic-ray emission, the contribution to the total cosmic-ray flux observed by a given observatory
can be written as,
\begin{equation}
I_{CR}(E,\hat{n}) = \frac{L_{CR}\ W_s(\hat{n})}{4 \pi D_s^2\ (1+z_s) \langle E \rangle_0 } \ K_{CR}\ P_{CR}(E),
\label{eq:CRFlux}
\end{equation}
where $z_s$ is the redshift of the source, $\langle E_{0} \rangle$ is the mean energy of
particles in the source, $L_{CR}$ is the cosmic-ray luminosity, $P_{CR}(E)$ is the energy distribution of particles arriving on Earth
and $K_{CR}$ is the number of cosmic rays arriving on Earth per injected particle. $K_{CR}$ and $P_{CR}(E)$ include the physics of the
propagation of the cosmic rays. $W_s$ is the weight of a specific point source to the total measured flux taking into account the
exposure of the observatory (see Ref.~\cite{bib:model} for details). Note that $W_s(\hat{n})$ depends on the location on the Earth's
surface of a given observatory and on the position on the sky of the source which is given by the unit vector $\hat{n}$.

For this source, a secondary gamma-ray flux is also generated as a consequence of the propagation of the cosmic rays in the intergalactic
medium, which is proportional to its cosmic-ray luminosity. In analogy to equation~(\ref{eq:CRFlux}) it is possible to write the gamma-ray
flux at Earth as a function of the cosmic-ray luminosity,
\begin{equation}
I_{\gamma}(E_\gamma) = \frac{L_{CR}}{4 \pi D_s^2\ (1+z_s) \langle E \rangle_0 } \ K_{\gamma}\ P_{\gamma}(E_\gamma),
\label{eq:GammaFlux}
\end{equation}
where $K_{\gamma}$ is the number of gamma-rays generated per injected cosmic-ray particle and $P_{\gamma}(E_\gamma)$ is the energy
distribution of the gamma-rays observed at Earth. $K_{\gamma}$ and $P_{\gamma}(E_\gamma)$ summarize the propagation of the particles
and production of GeV-TeV gamma-rays.

From equation (\ref{eq:CRFlux}) it can be seen that the normalization constant $C_s=L_{CR}\ W_s$ is limited by the observed cosmic-ray
spectrum (more precisely by the upper limit of the spectrum at a given Confidence Level (CL)), i.e.~an individual source cannot exceed the
total cosmic-ray flux observed by a given Observatory. Then, the maximum gamma-ray flux consistent with the cosmic-ray observations is given
by,
\begin{equation}
I_{\gamma}^{max}(E_\gamma,\hat{n}) = \frac{C_s^{max}}{W_s(\hat{n})\ 4 \pi D_s^2\ (1+z_s) \langle E \rangle_0 } \ K_{\gamma}%
\ P_{\gamma}(E_\gamma).
\label{eq:GammaFluxMax}
\end{equation}

Let us define the function $I^{UHECR}_\gamma(E_\gamma) \equiv W_s(\hat{n})\ I_{\gamma}^{max}(E_\gamma,\hat{n})$ which is independent of the
position of the source in the sky. If the integral of the gamma-ray flux above a given energy threshold is considered it takes the following
form,
\begin{equation}
I^{UHECR}_\gamma(> E_{th}) = \frac{C_s^{max}}{4 \pi D_s^2\ (1+z_s) \langle E \rangle_0 } \ K_{\gamma}\ %
\int_{E_{th}}^\infty dE_\gamma \ P_{\gamma}(E_\gamma).
\label{eq:GammaFluxMaxInt}
\end{equation}
Therefore, if an upper limit on the gamma-ray flux coming from gamma-ray observations, multiplied by $W_s$, is smaller than $I^{UHECR}_\gamma$,
it means that the gamma-ray observations are restrictive enough to set an upper limit on the cosmic-ray luminosity. In the other case, the
gamma-ray observations cannot give a valid upper limit on the cosmic-ray luminosity. Note that $I^{UHECR}_\gamma$ is a very useful parameter
when several sources are analyzed.

In order to calculate $I_\gamma^{UHECR}$ the program CRPropa~\cite{bib:crpropa} has been used to simulate sources located at a fixed distance
from Earth varying from 10 to 200 Mpc in steps of 10 Mpc. For each distance, $10^7$ particles have been simulated in the one-dimensional
approximation. The UHECR particles have been considered to be protons and iron nuclei emitted following a power law with an exponential cutoff
spectrum,
\begin{equation}
\frac{dN}{dE dt}=\frac{L_{CR}}{C_0}\ E^{-\alpha} \exp(-E/E_{cut}),
\end{equation}
where the spectral index ($\alpha$) varies form 2 to 2.8 in steps of 0.1 and the cutoff energy ($E_{cut}$) varies from $Z \times 10^{20}$ to
$Z \times 10^{21}$ eV where $Z$ is the nucleus charge. Here $C_0$ is a normalization constant. The propagation takes into account all energy
losses. Pair production is considered as a continuous energy loss. Nuclei undergo photo-disintegration and the generator SOPHIA~\cite{bib:sophia}
is used to simulate pion production. The propagation of UHECR generates the flux of GeV-TeV gamma-rays at Earth for each considered source. The
simulated UHECR flux is normalized to the flux measured by a given UHECR experiment. This normalization yields the corresponding normalization
of the secondary TeV-GeV gamma-rays flux.

\begin{figure}
\centering
\includegraphics[angle=90,width=0.8\textwidth]{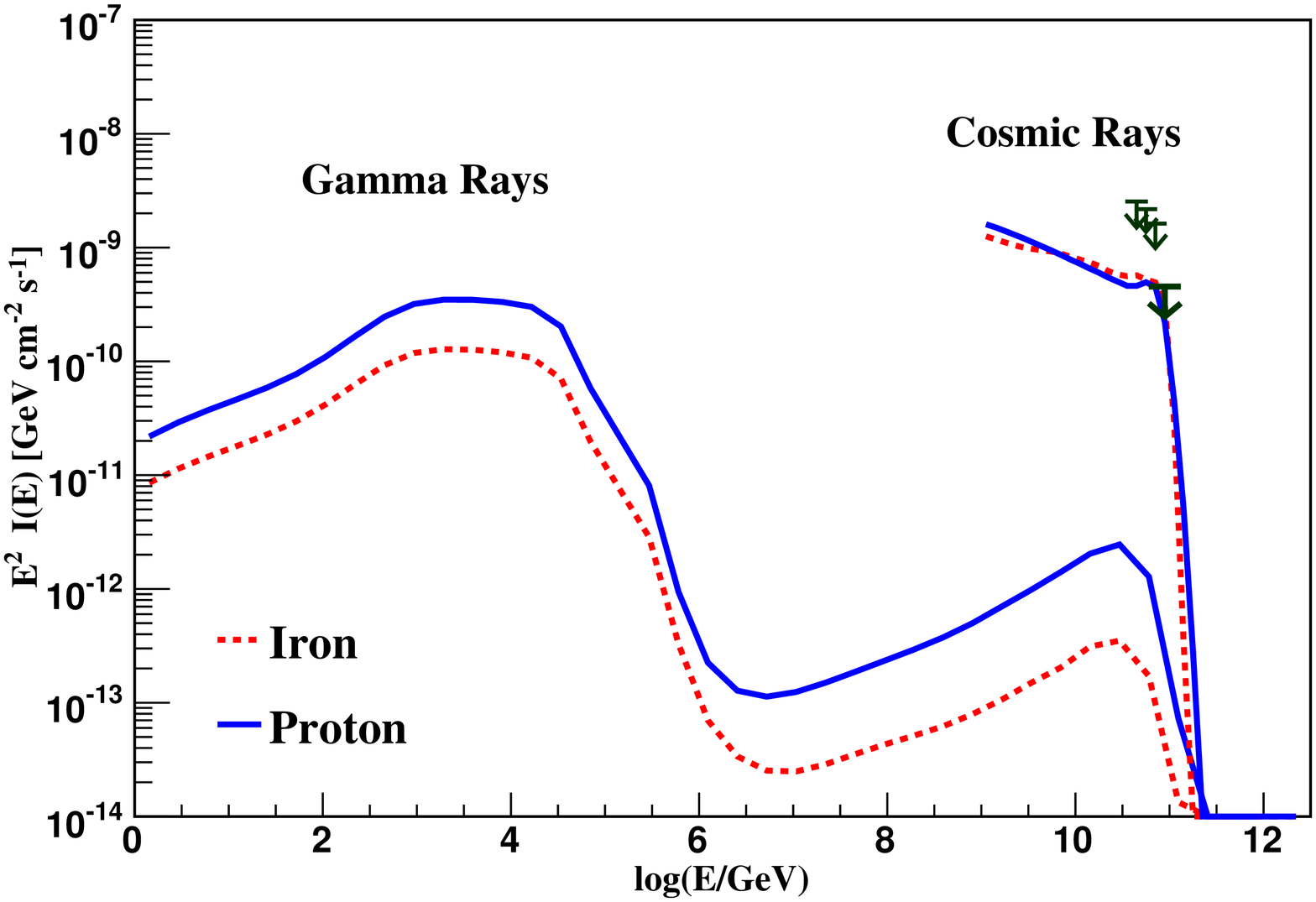}
\caption{Cosmic-ray and gamma-ray spectra simulated with CRPropa~\cite{bib:crpropa}. The source is 90 Mpc away from Earth, the injected cosmic ray
spectrum corresponds to a power law with an exponential cutoff such that $\alpha = 2.4$ and $E_{cut} = Z \times 10^{20.5}$ (see section~\ref{sec:source:sel}).
The cosmic-ray flux arriving on Earth was normalized to the flux measured by the Pierre Auger Observatory. The corresponding normalization was applied to the
gamma-ray flux as well. Protons and iron nuclei were considered as primary particles.}
\label{fig:spectra:gamma:uhecr}
\end{figure}

Figure~\ref{fig:spectra:gamma:uhecr} shows an example of the discussed procedure for a source 90 Mpc away from Earth. The simulated UHECR
spectrum was normalized considering the upper limit on the energy spectrum measured by the Pierre Auger Observatory at 95\% CL. The arrows
in figure~\ref{fig:spectra:gamma:uhecr} show the upper limit of the flux measured by the Pierre Auger Observatory used to normalize the simulated
cosmic-ray spectrum (see Ref.~\cite{bib:model} for details on how the upper limit on the spectrum, at a given CL, is calculated from the
published data of the Pierre Auger and Telescope Array observatories). It is clear that only one point measured by the Pierre Auger Observatory
limits the cosmic-ray flux. Following this procedure the normalization of the secondary GeV-TeV gamma-ray spectrum is obtained.

\begin{figure}
  \centering
  \subfloat[$E_{cut}$ study.]{\includegraphics[angle=90,width=0.45\textwidth]{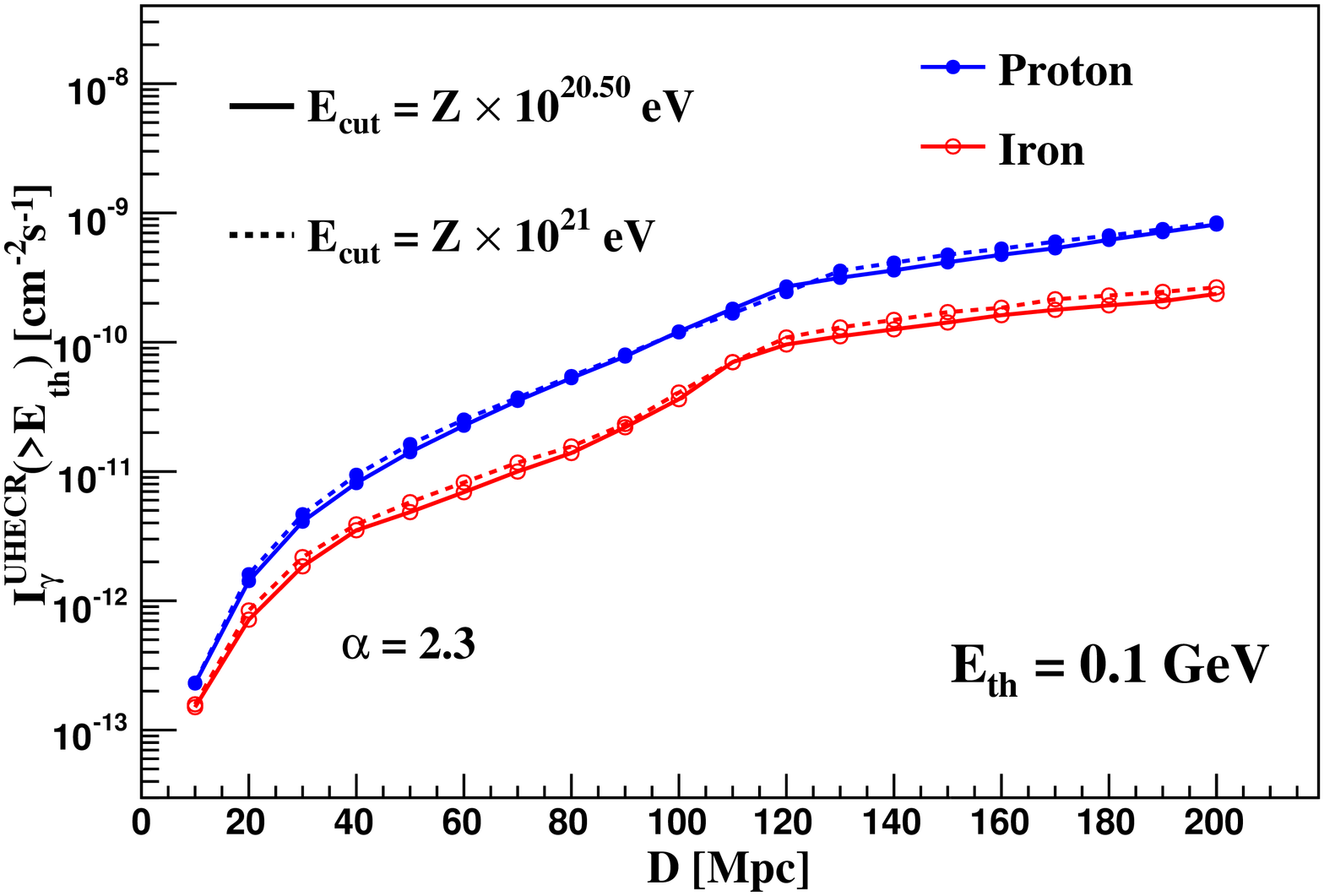}}
  \subfloat[$\gamma$ study.]{\includegraphics[angle=90,width=0.45\textwidth]{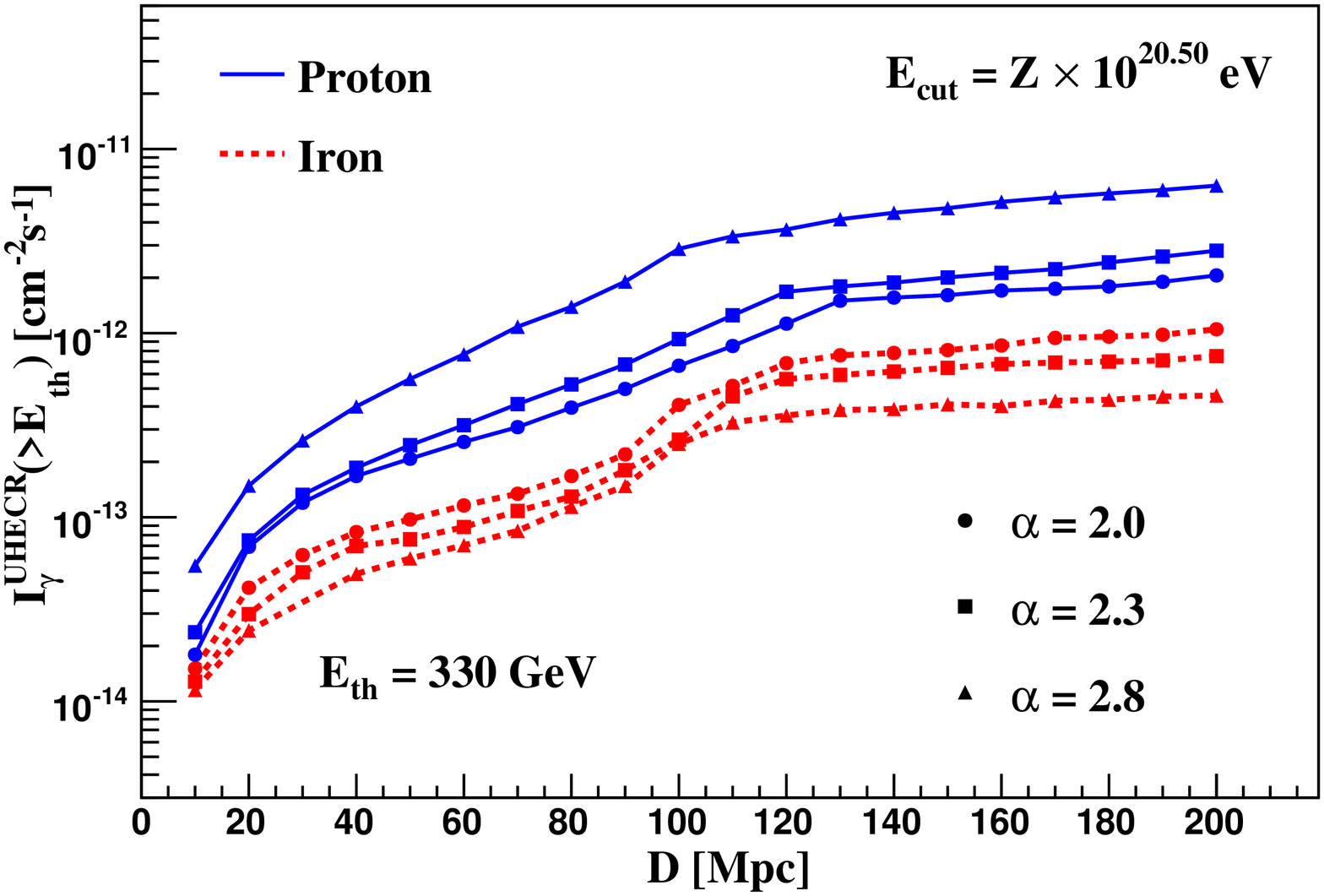}}\\
  \subfloat[$E_{th}$ study.]{\includegraphics[angle=90,width=0.45\textwidth]{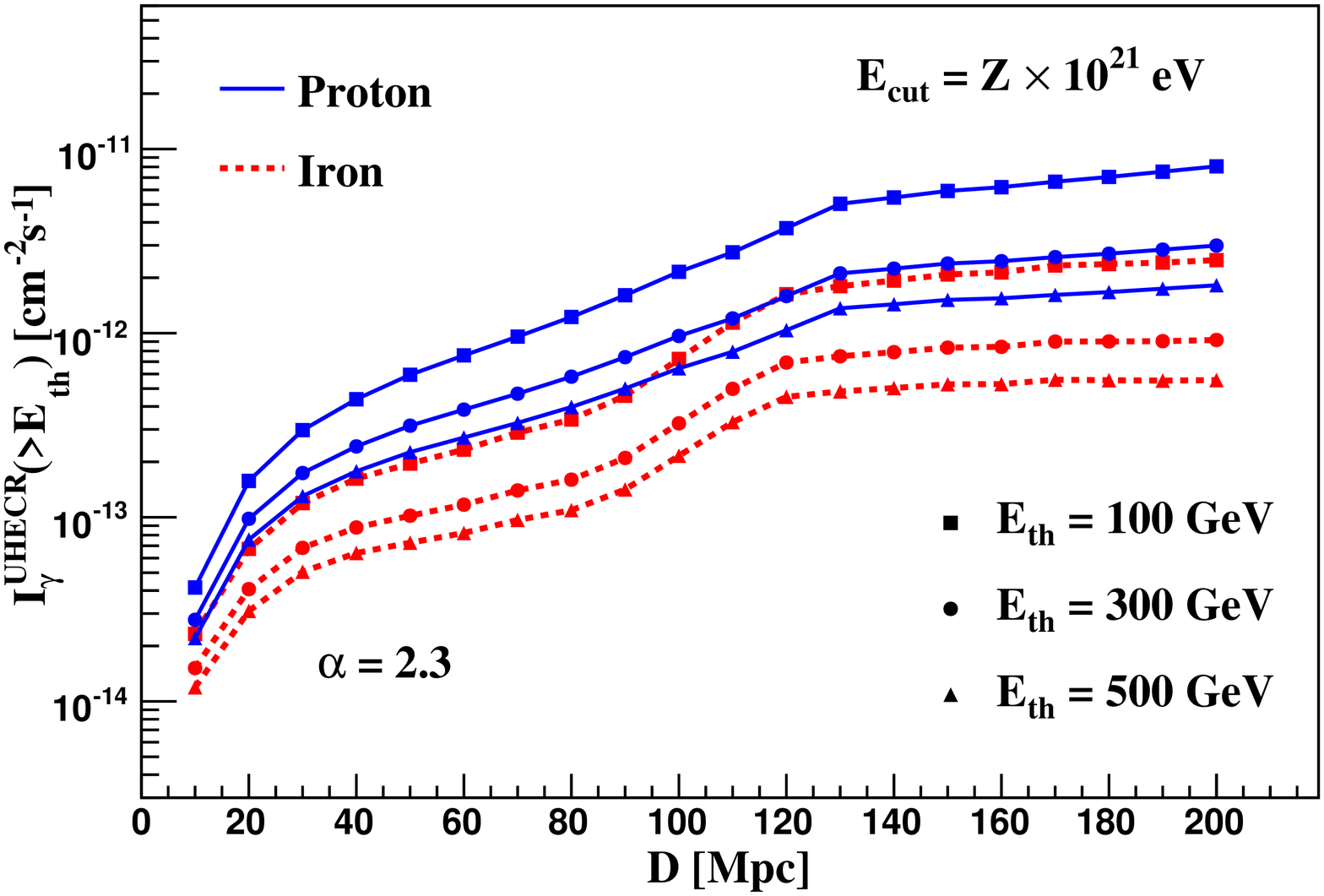}}\\
\caption{The plots show $I_\gamma^{UHECR}$ as a function of the source distance, calculated by using the upper limit on the flux observed by
the Pierre Auger Observatory at 95 \% CL. Figure (a) shows the dependence of $I_\gamma^{UHECR}$ on $E_{cut}$ for fixed $\alpha =2.3$ and fixed
$E_{th} = 0.1$ GeV. Figure (b) shows the dependence of $I_\gamma^{UHECR}$ on the spectral index $\alpha$ for fixed
$E_{cut} = Z \times 10^{20.5}$ eV and fixed $E_{th} = 330$ GeV. Figure (c) shows the dependence of $I_\gamma^{UHECR}$ on the energy threshold
($E_{th}$) for fixed $E_{cut} = Z \times 10^{21}$ eV and fixed $\alpha = 2.3$.}
\label{bib:upper:limits:distance}
\end{figure}

This procedure was repeated for all simulated sources as a function of distance. Figure~\ref{bib:upper:limits:distance} shows
$I^{UHECR}_{\gamma}(> E_{th})$ for different parameters of the injection spectrum considered in this work. Figure~\ref{bib:upper:limits:distance}a
shows $I^{UHECR}_{\gamma}(> E_{th})$ obtained by using the upper limit flux at 95\% CL. of the Pierre Auger Observatory, for two different values
of $E_{cut}$ and for protons and iron nuclei as primaries. It is clear that the cutoff energy of the source has a marginal effect on
$I^{UHECR}_{\gamma}$. Figure~\ref{bib:upper:limits:distance}b shows the same study but for different values of the spectral index ($\alpha$). As
expected the influence of $\alpha$ on the determination of $I^{UHECR}_{\gamma}$ is larger. Note the different dependence of $I^{UHECR}_{\gamma}$
on $\alpha$ for proton and iron primaries. Finally, figure~\ref{bib:upper:limits:distance}c shows the variation of $I^{UHECR}_{\gamma}(> E_{th})$
for different energy thresholds. The $E_{th}$ of the gamma-ray measurement has a strong influence on $I^{UHECR}_{\gamma}$. However, this value is
always determined in each gamma-ray measurement and therefore it has to be taken into account in each particular case.

\begin{figure}
  \centering
  \subfloat[FERMI-LAT - TA - $E_{th} = 0.1 GeV$]{\includegraphics[angle=90,width=0.45\textwidth]{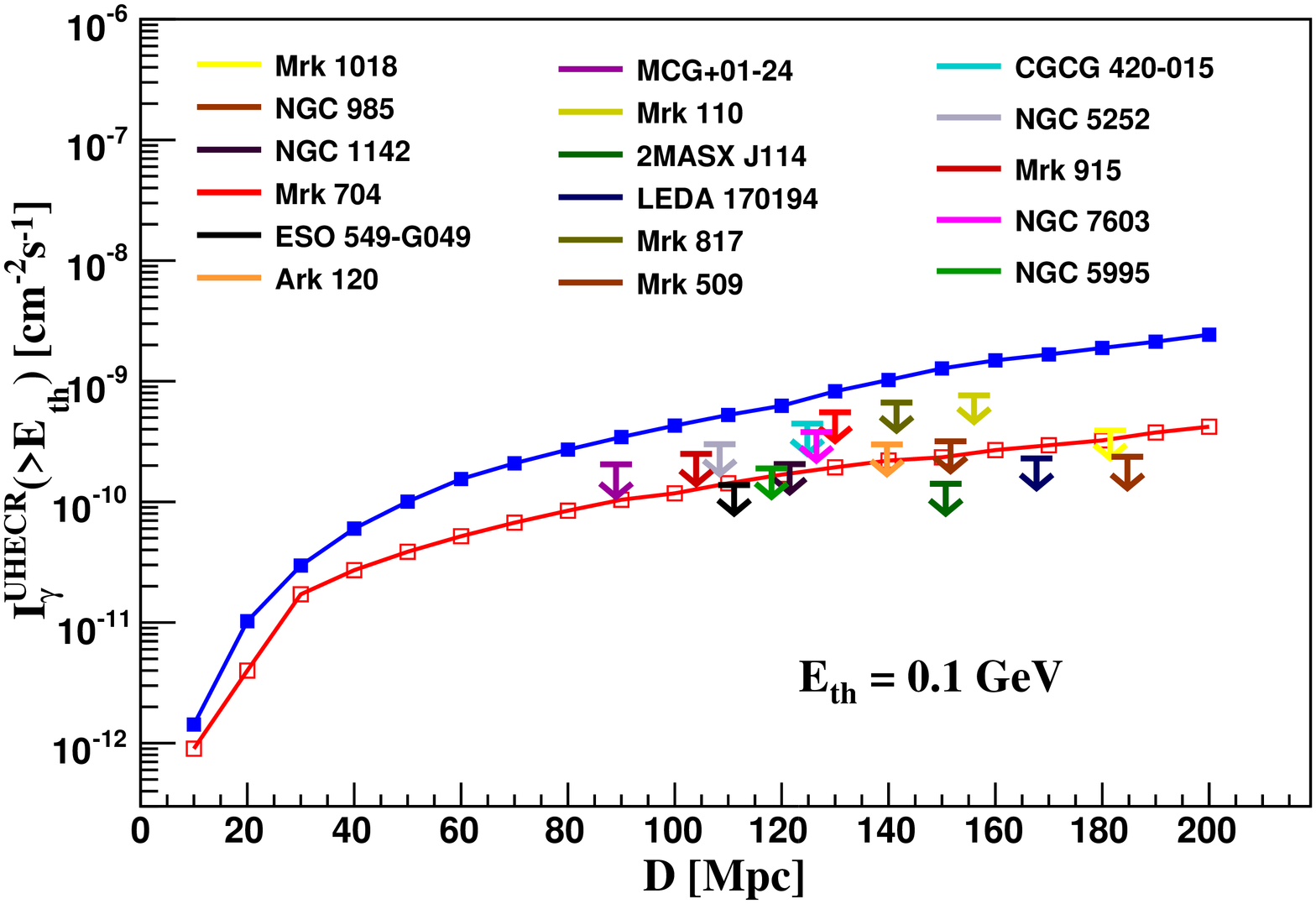}}
  \subfloat[FERMI-LAT - Auger - $E_{th} = 0.1 GeV$]{\includegraphics[angle=90,width=0.45\textwidth]{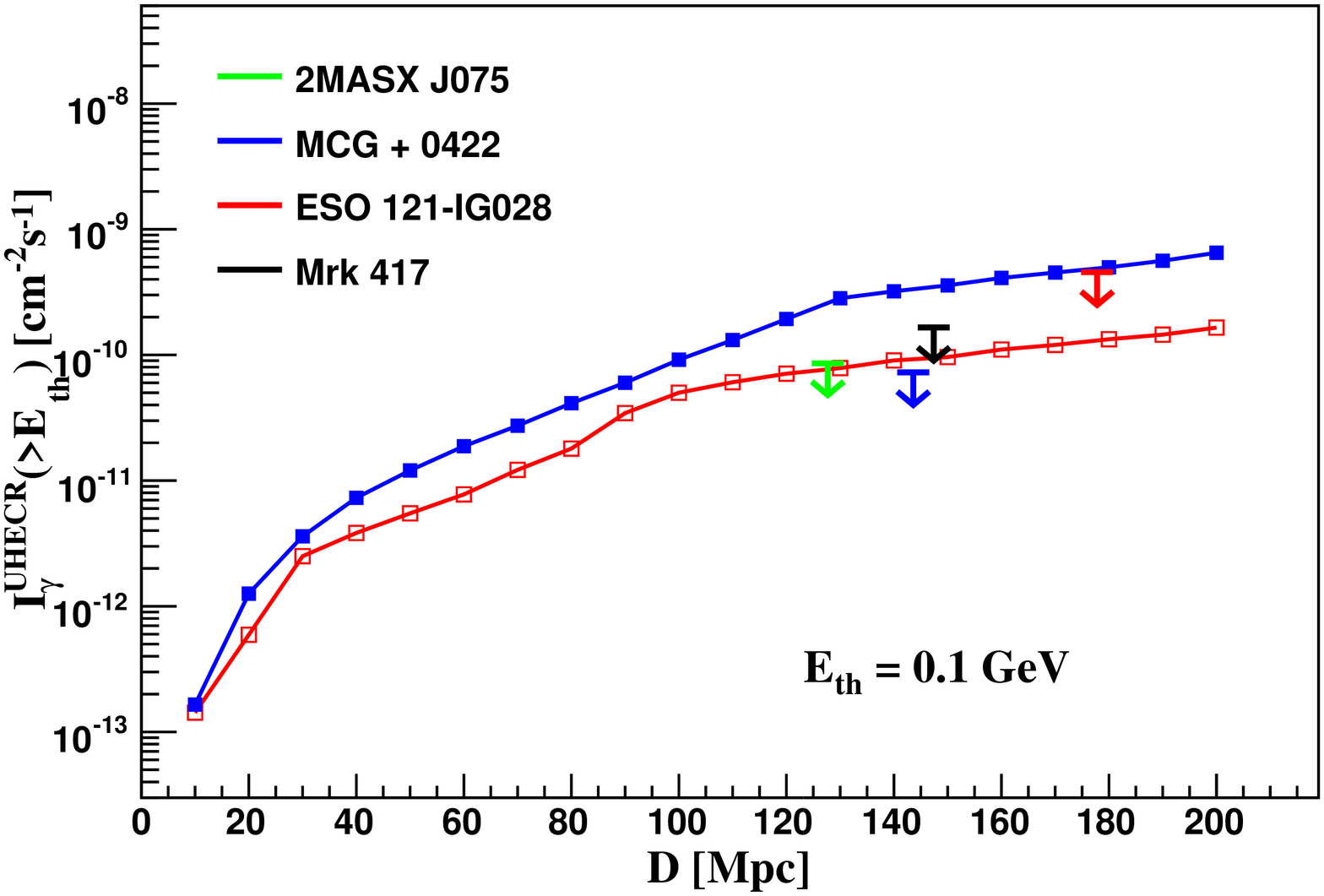}}\\
  \subfloat[FERMI-LAT - TA - $10 < E_\gamma < 100 GeV$]{\includegraphics[angle=90,width=0.45\textwidth]{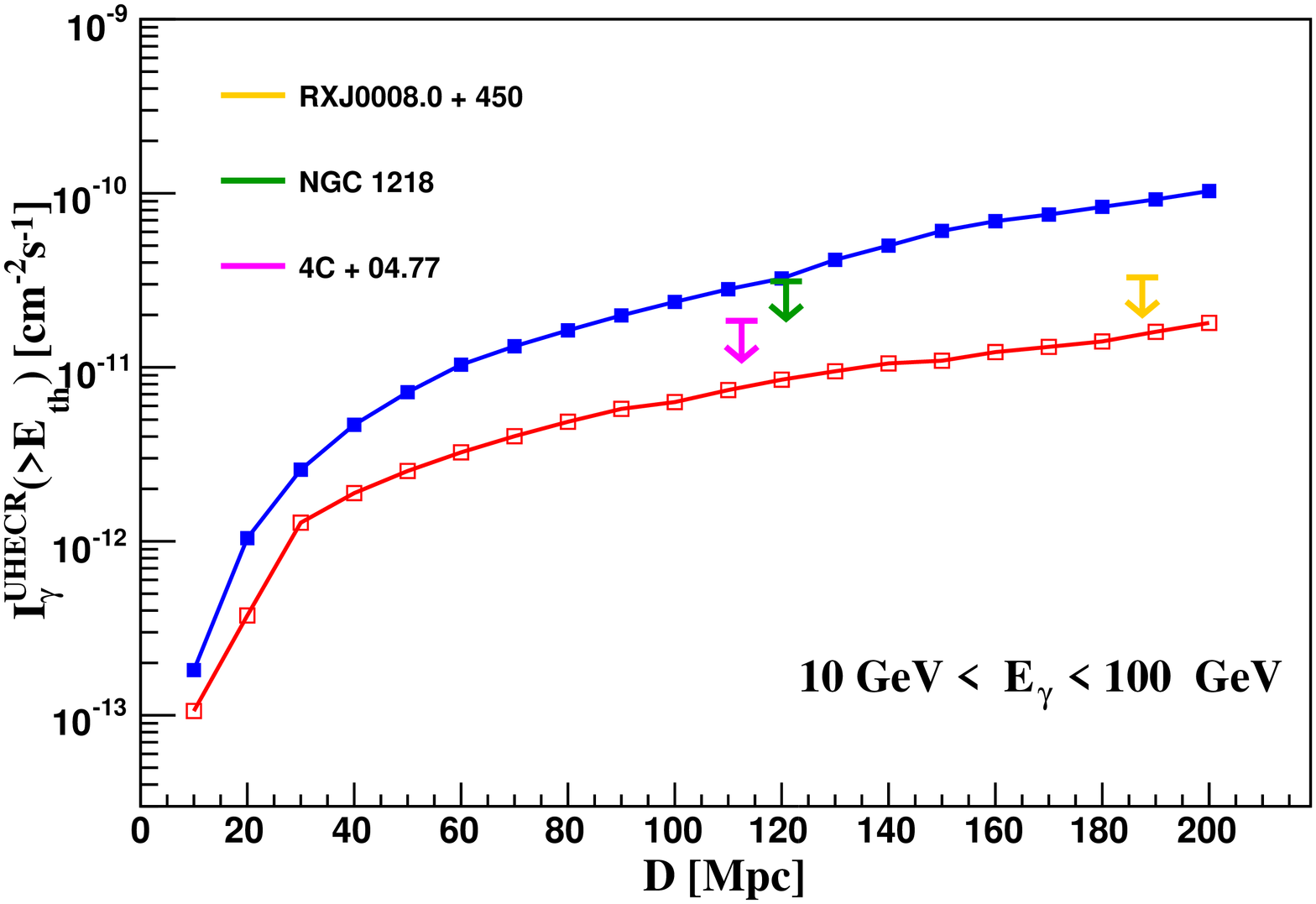}}\\
  \caption{Upper limit on the integral flux of gamma-rays ($I^{UHECR}_{\gamma}$ at 95\% CL) as a function of distance of the
source. The red lines with open squares correspond to iron primaries. The blue lines with closed squares correspond to primary protons. Arrows are
the measured gamma-ray data (multiplied by the corresponding $W_s$). For all plots the injected spectrum was considered to be a power law with
$\alpha = 2.4$ and $E_{cut} = Z \times 10^{10.5}$ eV. In each plot the caption shows: The observatory which obtained the gamma-ray data (FERMI-LAT),
the UHECR observatory (Auger or TA), and the energy of the gamma-rays used to calculate the integral flux.}
  \label{fig:upper:limits:select}
\end{figure}

Figure~\ref{fig:upper:limits:select} shows the comparison of $I^{UHECR}_{\gamma}$ (lines) with the measured upper limits (multiplied by the
corresponding $W_s$) obtained from different gamma-ray experiments. As mentioned before, the UHECR spectra used in this paper have been measured
by the Pierre Auger~\cite{bib:auger:spectrum} and Telescope Array~\cite{bib:ta:spectrum} (TA) observatories. The use of the Pierre Auger or the
Telescope Array data depends on the exposure of each observatory in the direction of the source.

Figure~\ref{fig:upper:limits:select} is used to select the sources for which the gamma-ray measurement is restrictive enough in order to
limit the UHECR luminosity. Since an upper limit is the aim of the calculation, the worst case scenario:  $\alpha = 2.0$ and
$E_{cut} = 10^{20.5}$ eV for proton primaries, and $\alpha = 2.8$ and $E_{cut} = 26 \times 10^{20.5}$ eV for iron nuclei, is used.

The selection criteria (figure ~\ref{fig:upper:limits:select}) have to be evaluated for each $E_{th}$ and UHECR experiment.
Figure~\ref{fig:upper:limits:select} shows some examples for which it was possible to show more than one source in the same figure. Not all
sources in~\cref{tab:1,tab:2,tab:3} are shown. If the measured upper limit of one source is below the blue line (closed circles) then the
measured upper limit on the integral flux of gamma-rays is restrictive enough to set an upper limit on the proton luminosity. If the measured
upper limit of a source is below the red line (open squares) then the measured upper limit on the integral flux of gamma-rays is restrictive
enough to set an upper limit on the total UHECR luminosity.

The gamma-ray experiments have observed a few thousands of sources closer than 200 Mpc. Using plots similar to the ones in
figure~\ref{fig:upper:limits:select} it was possible to select the twenty-eight sources for which the upper limit on the integral flux of
gamma-rays is restrictive enough in order to set an upper limit on the UHECR proton or even total luminosity.

\section{Upper limits on the UHECR luminosity}
\label{sec:upper}

The selection of the sources presented in section~\ref{sec:source:sel} is based on the UHECR measured spectrum from which an upper limit on the
integral of the gamma-ray flux is calculated. The relationship between the UHECR measured spectrum and the calculated upper limit on the integral
of the gamma-ray flux is obtained by using propagation models. Regarding the selected sources, from the measured upper limit on the gamma-ray
flux of a given source and using propagation models, an upper limit on the UHECR flux can be obtained.

\begin{figure}
  \centering
  \includegraphics[angle=90,width=0.40\textwidth]{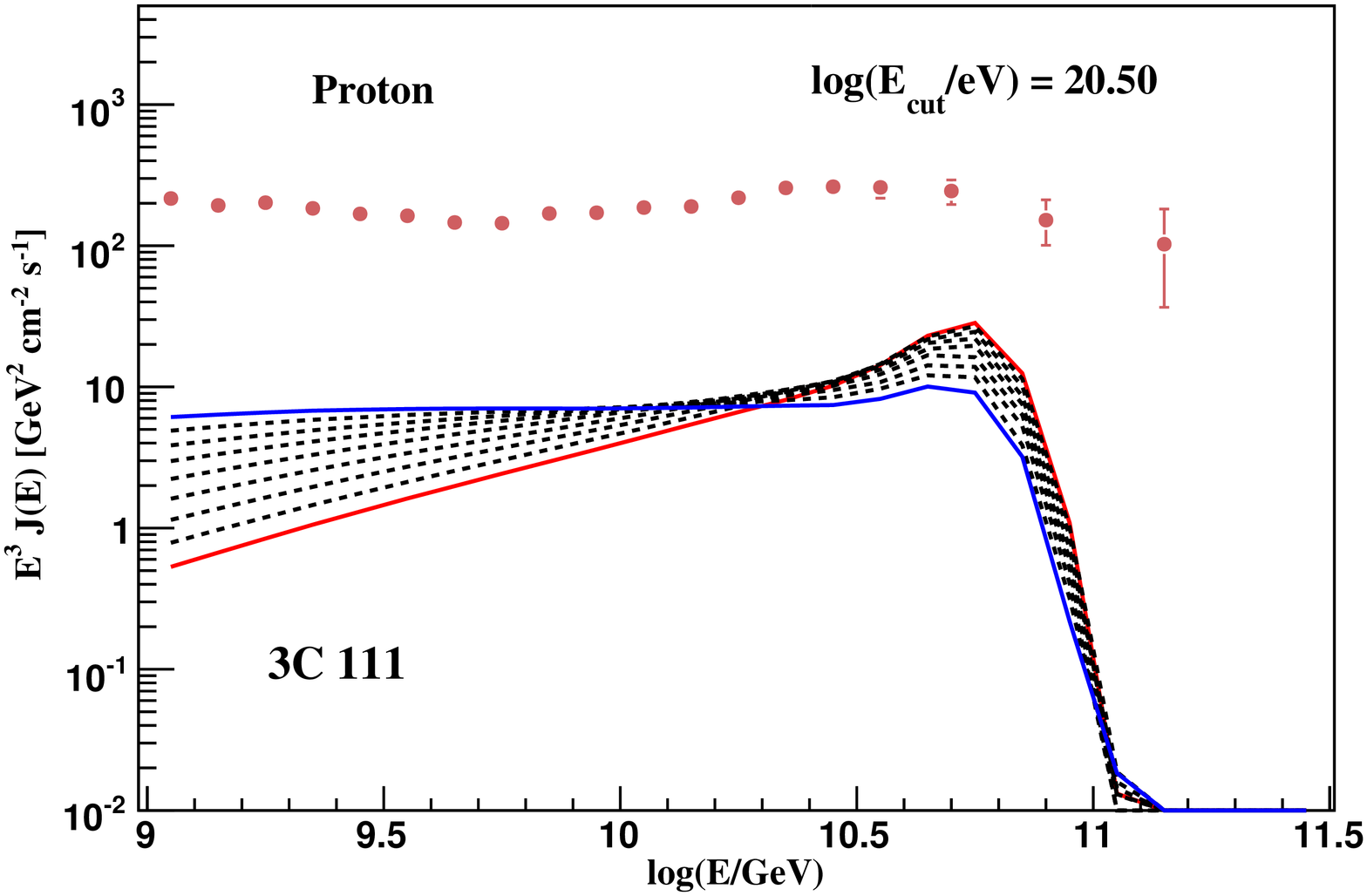}
  \includegraphics[angle=90,width=0.40\textwidth]{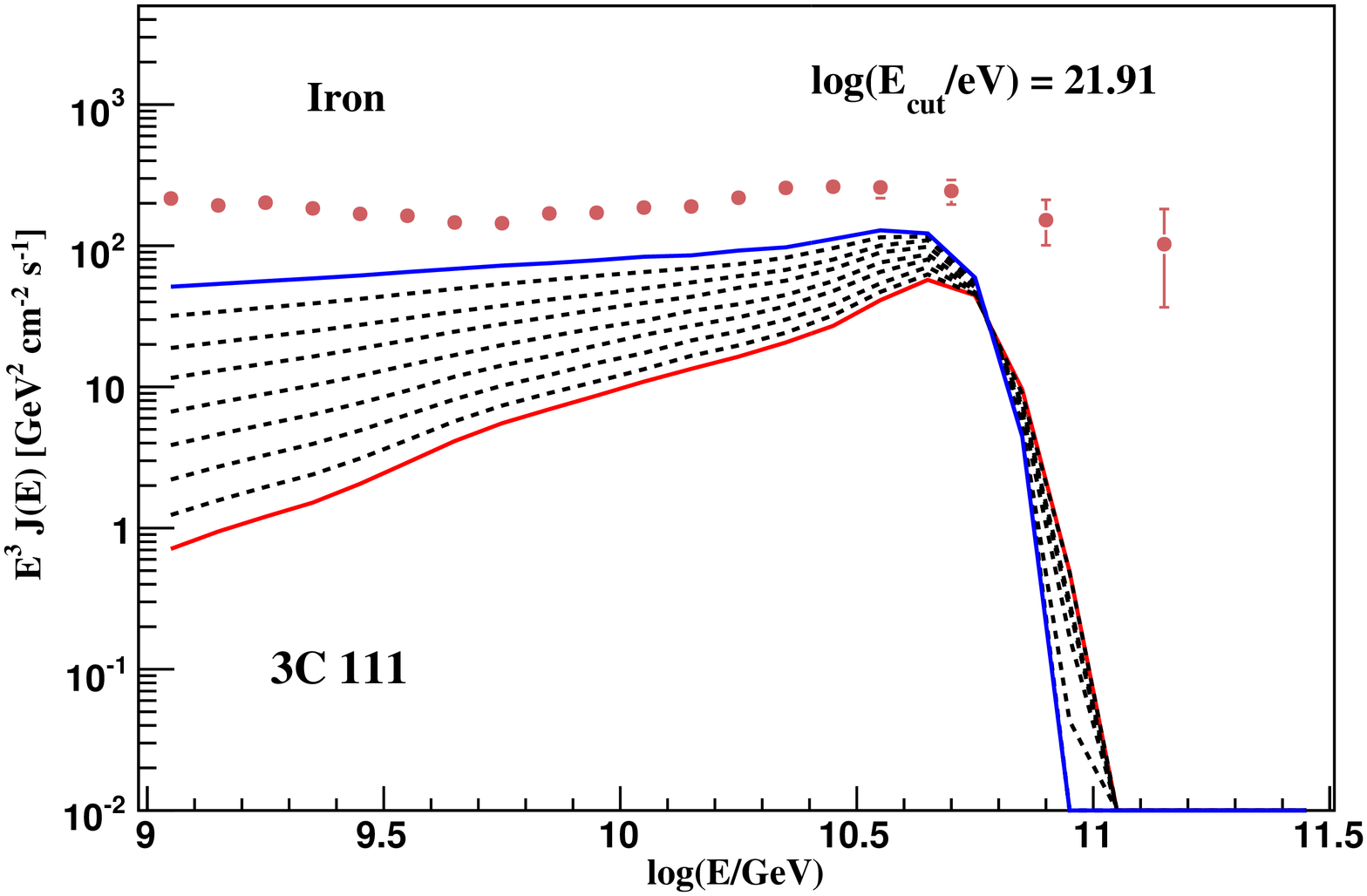}
  \includegraphics[angle=90,width=0.40\textwidth]{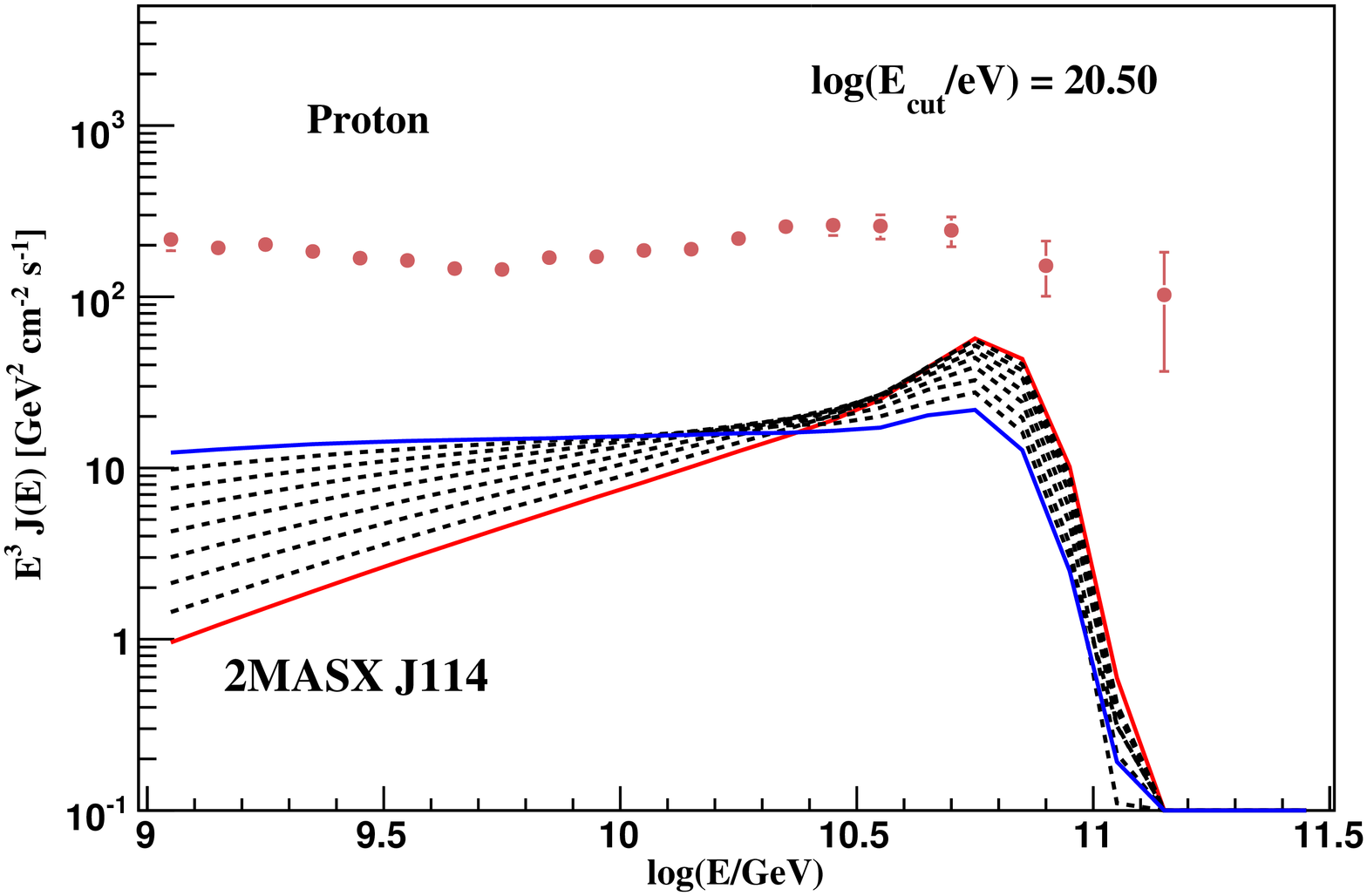}
  \includegraphics[angle=90,width=0.40\textwidth]{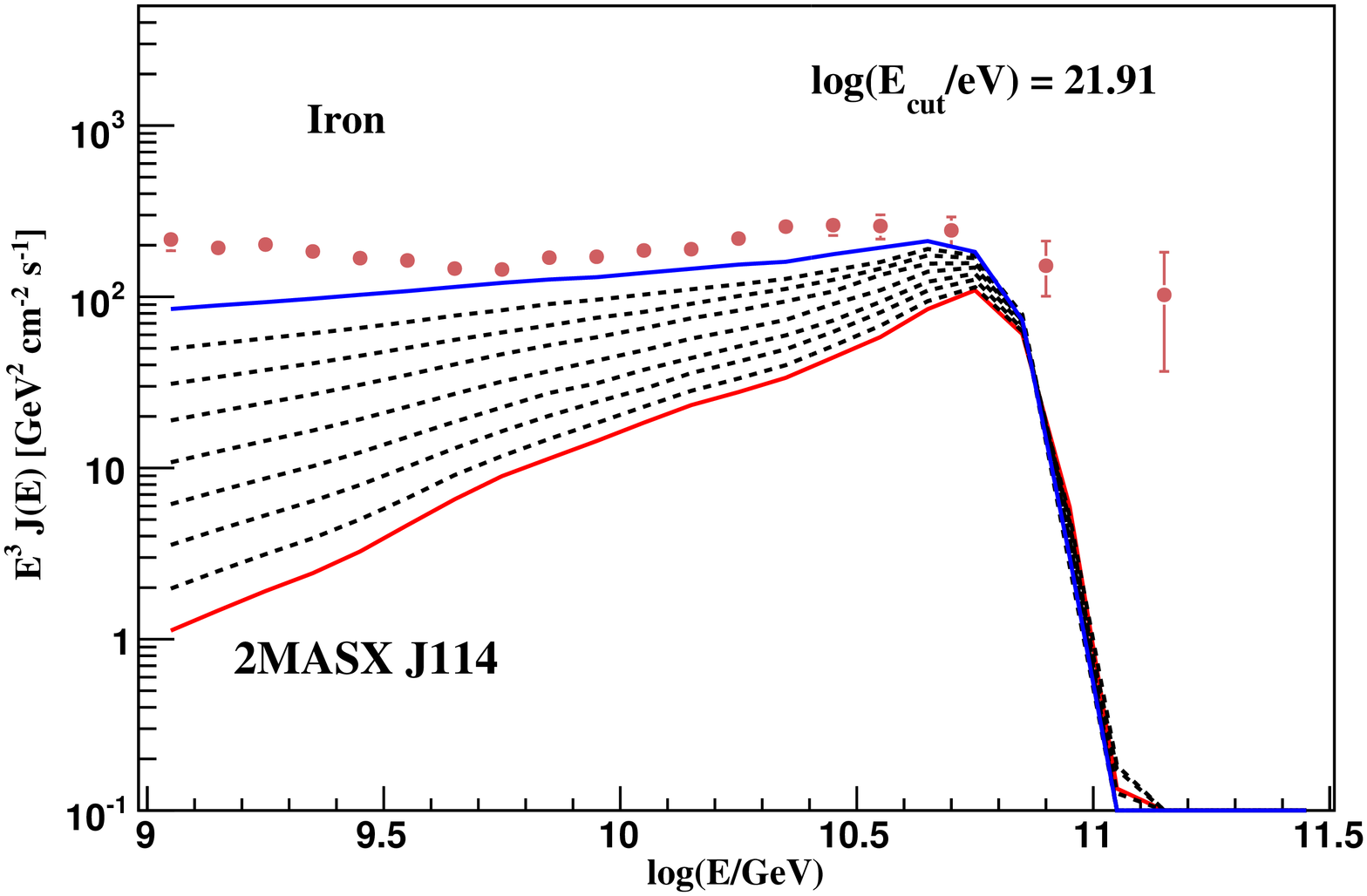}
  \includegraphics[angle=90,width=0.40\textwidth]{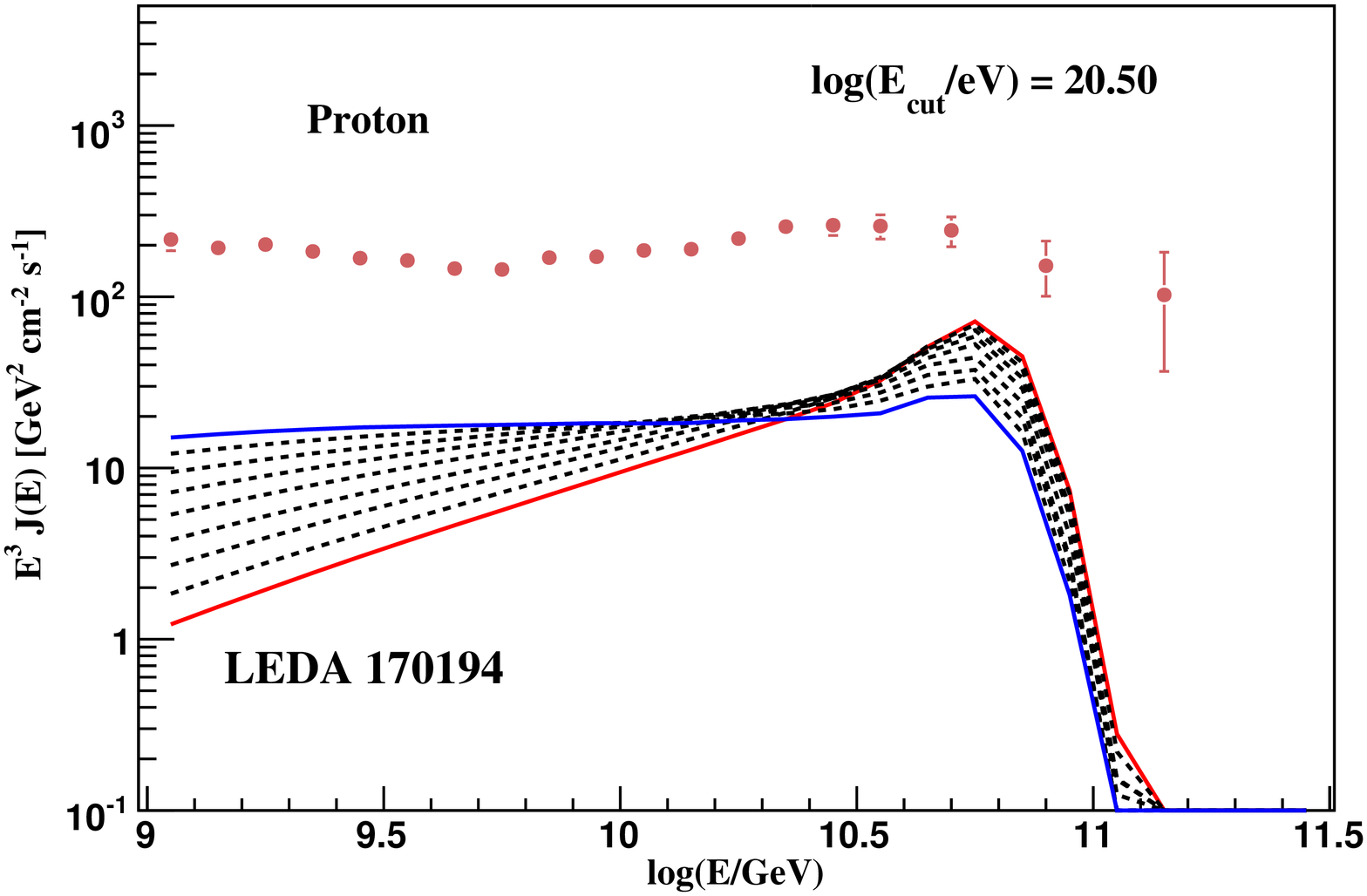}
  \includegraphics[angle=90,width=0.40\textwidth]{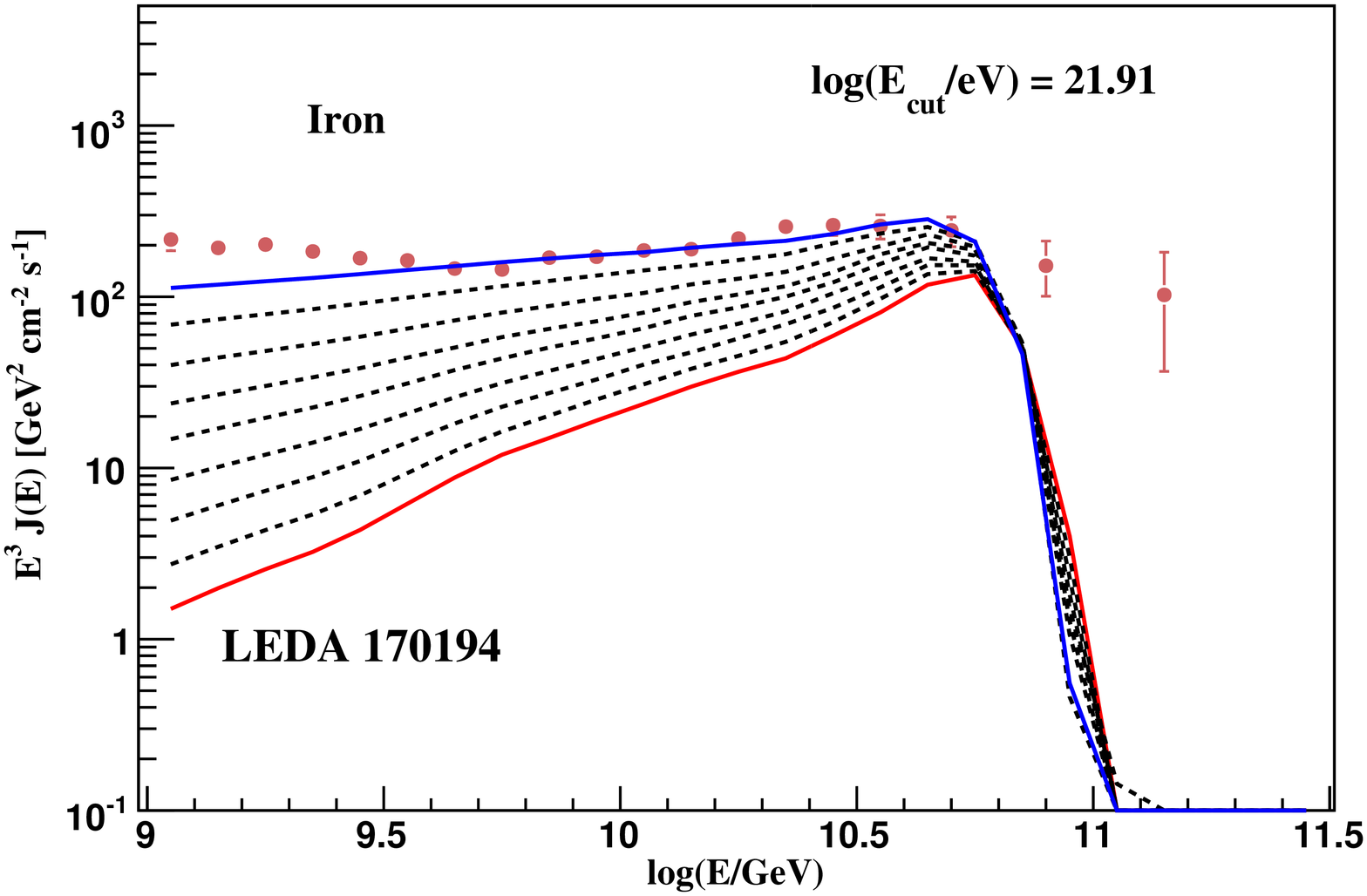}
  \includegraphics[angle=90,width=0.40\textwidth]{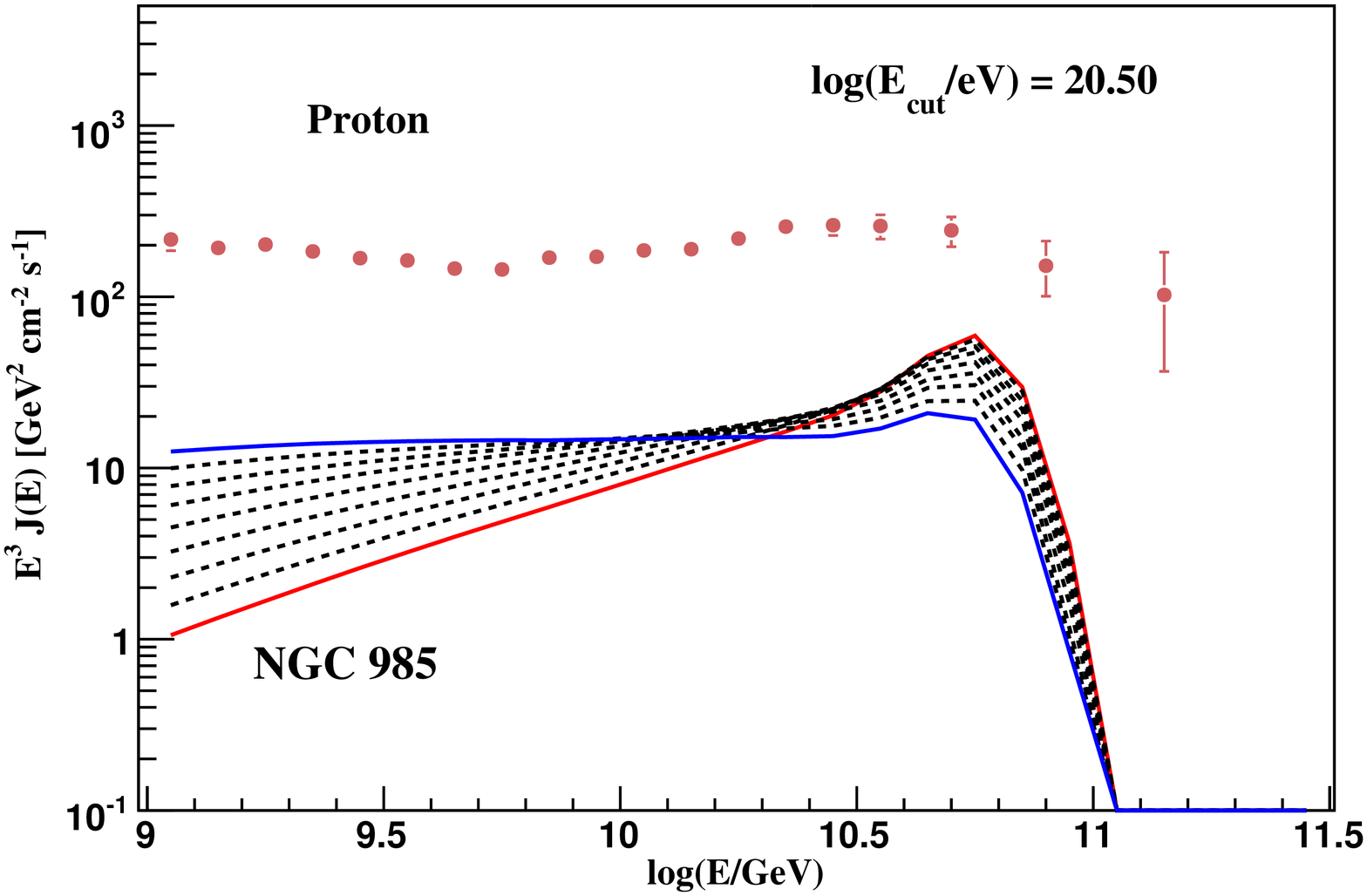}
  \includegraphics[angle=90,width=0.40\textwidth]{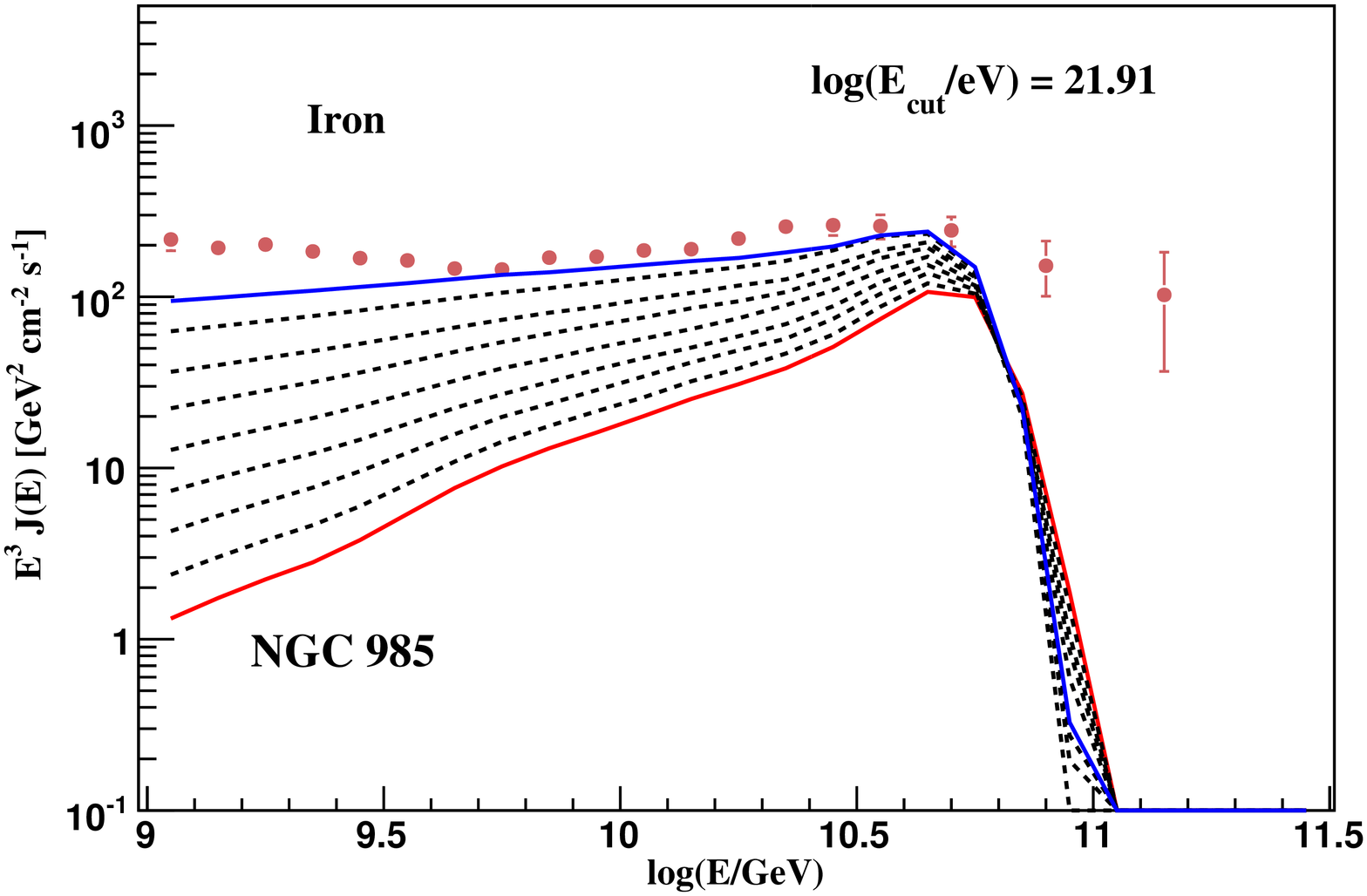}
  \includegraphics[angle=90,width=0.40\textwidth]{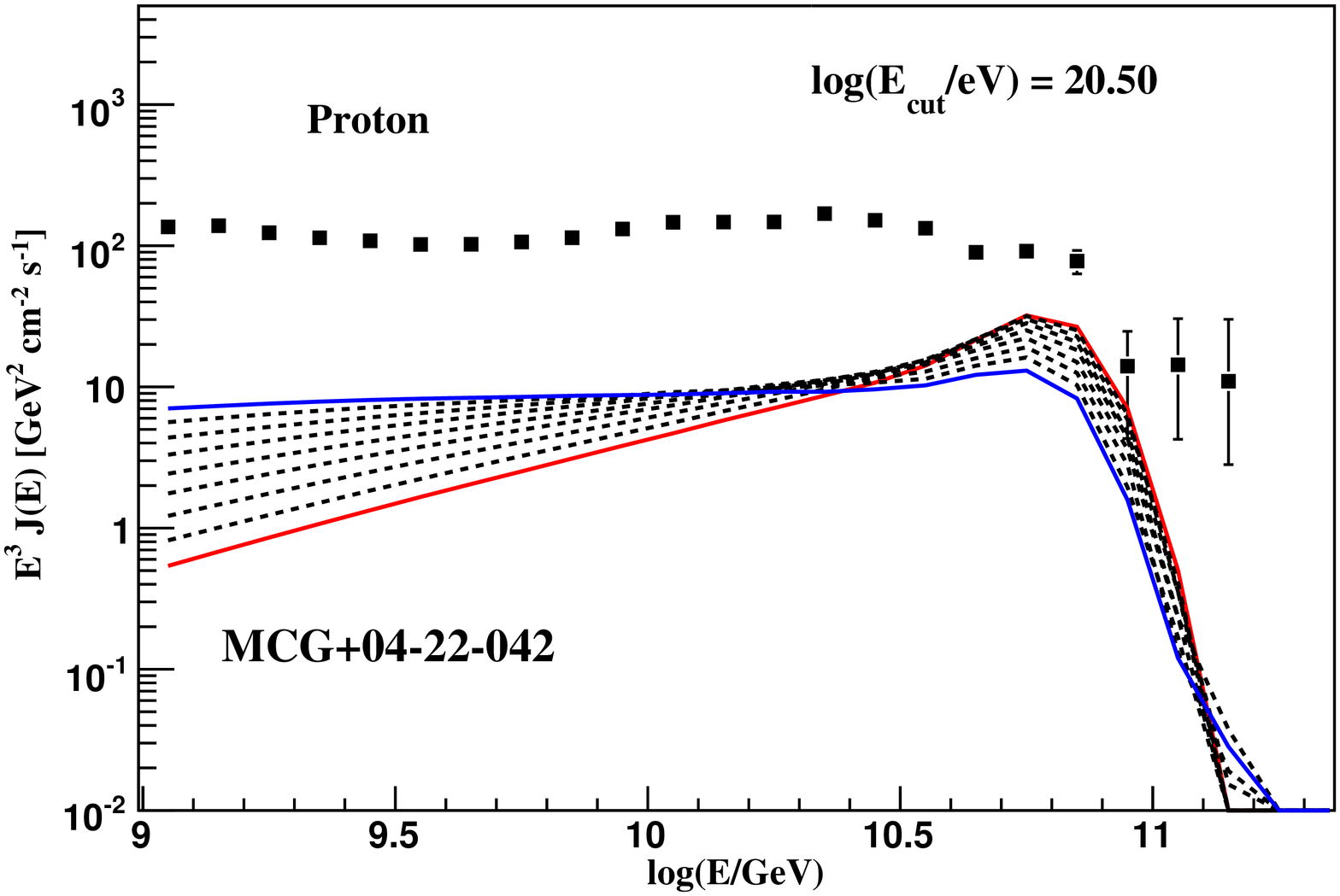}
  \includegraphics[angle=90,width=0.40\textwidth]{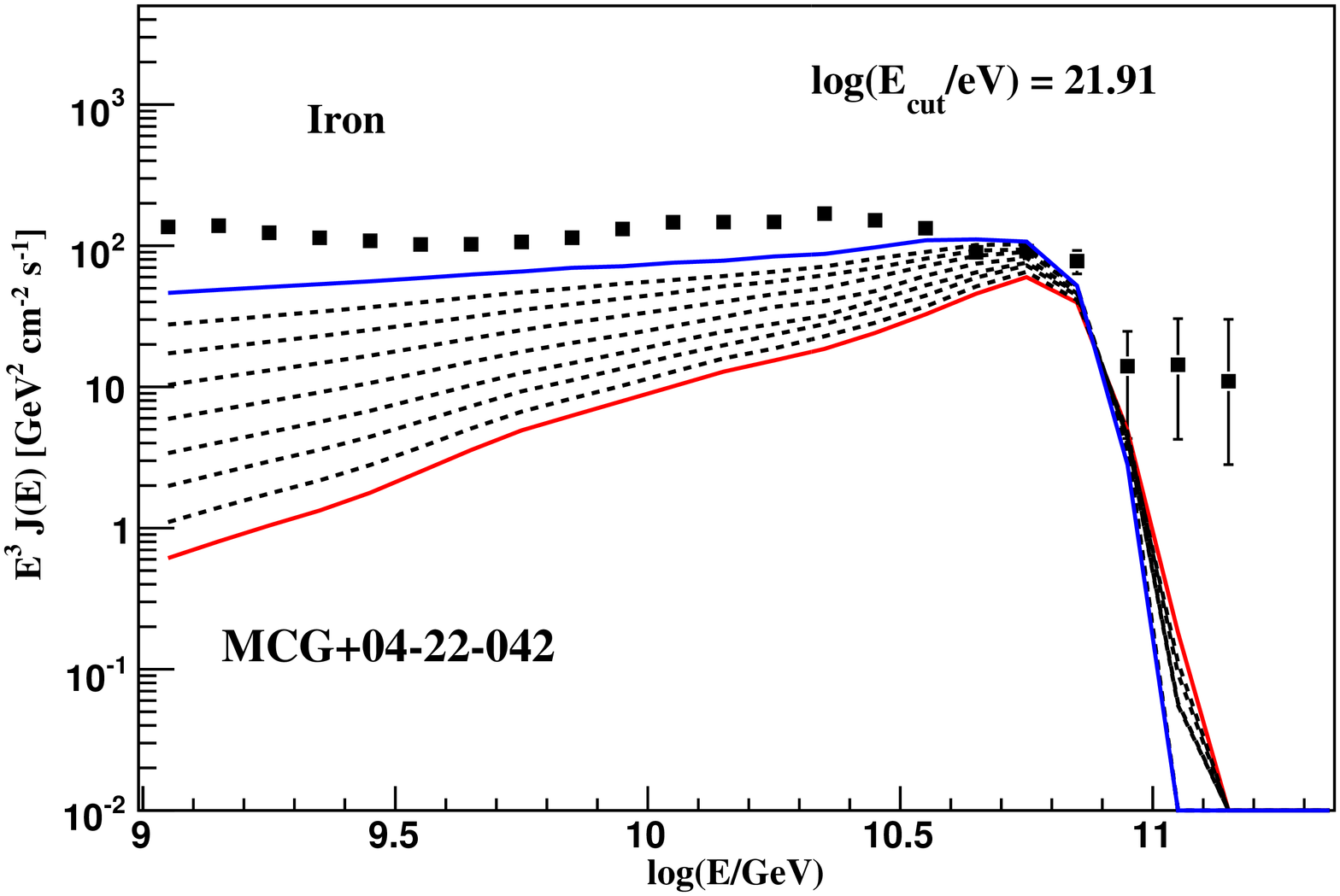}
  \caption{The lines show the cosmic-ray spectrum of the sources 3C 111, J11454045-1827149, LEDA 170194, NGC 985, and MCG+04-22-042 as calculated by using
the measured upper limit on the integral flux of GeV-TeV gamma-rays. For each source, nine spectra are shown with spectral index ($\alpha$) going from 2
(red) to 2.8 (blue) in steps of 0.1. Left: Primary proton. Right: Primary iron nucleus. The points with error bars correspond to the measured cosmic-ray
flux obtained by TA (top five) or Auger (last).}
  \label{fig:spectrum}
\end{figure}

Figure~\ref{fig:spectrum} shows the calculated UHECR spectrum of some selected sources based on the measured upper limit on the integral flux of
gamma-rays. The sources 3C 111, J11454045-1827149, LEDA 170194, and NGC 985 are best restricted by the Telescope Array measurements, therefore their spectra are shown
in comparison to the measured spectrum of the Telescope Array Observatory. The source MCG+04-22-042 is best restricted by the Pierre Auger Observatory data
and then the spectra are shown in comparison to the spectrum measured by the Pierre Auger Observatory. Left column of that figure shows the results
obtained for proton primaries and right column shows the results obtained for iron nuclei. The cutoff energy is such that
$E_{cut}^{Z} = Z  \times E_{cut}^{pr}$.

Figure~\ref{fig:spectrum} illustrates how the selection procedure described in section~\ref{sec:source:sel} selects sources for
which the upper limit on the UHECR flux is below the flux measured by UHECR experiments. It is also clear from this figure that within the
considered range of primary particles (proton to iron nuclei), the maximum flux is reached by iron nuclei. In other words, protons propagating
in the intergalactic medium generate more GeV-TeV gamma-rays than iron nuclei, therefore, a larger flux of iron nuclei is needed to generated
the measured upper limit on the integral flux of gamma-rays. Since iron nuclei are the primaries which generate the minimum flux of GeV-TeV
gamma-rays, an upper limit on the luminosity of iron nuclei is indeed a conservative upper limit to the total luminosity of the
source (see appendix~\ref{sec:iron:max} for details).

Given an upper limit on the integral gamma-ray flux, an upper limit on the cosmic-ray luminosity of the source can be obtained by using equation
(\ref{eq:GammaFlux}),
\begin{equation}
L_{CR}^{UL} = \frac{4\pi D^{2}_s(1+z_s)\langle E \rangle_{0}}{ \ K_{\gamma} {\mathop{\displaystyle
\int_{E_{th}}^{\infty} dE_\gamma\ P_{\gamma}(E_{\gamma})}}}\ I_{\gamma}^{UL}(> E^{th}_{\gamma}),
\label{eq:CRUL}
\end{equation}
where $I_{\gamma}^{UL}(> E^{th}_{\gamma})$ is the upper limit on the integral gamma-ray flux for a given confidence level and energy
threshold.

\begin{figure}
  \centering
  \includegraphics[angle=90,width=0.42\textwidth]{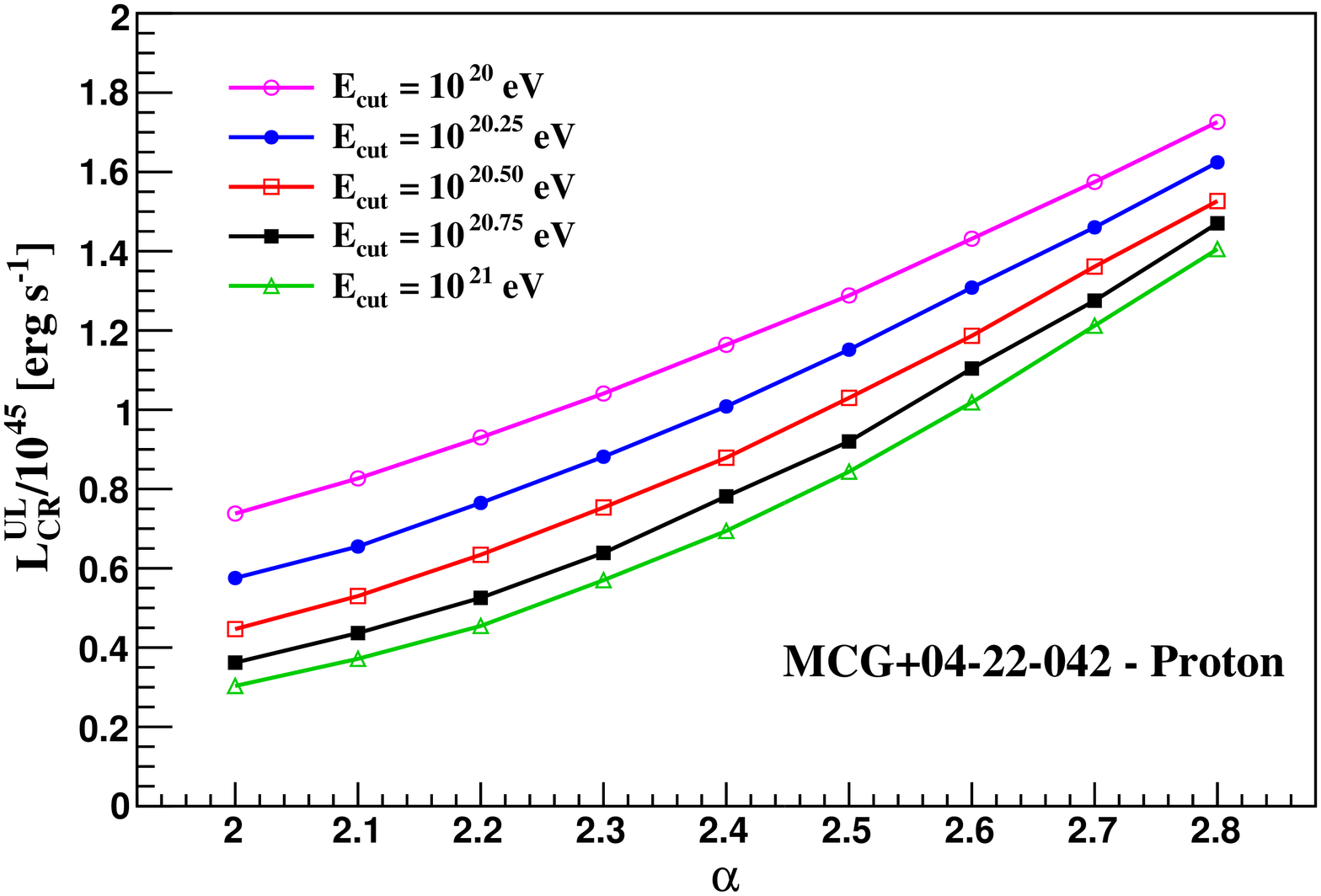}
  \includegraphics[angle=90,width=0.42\textwidth]{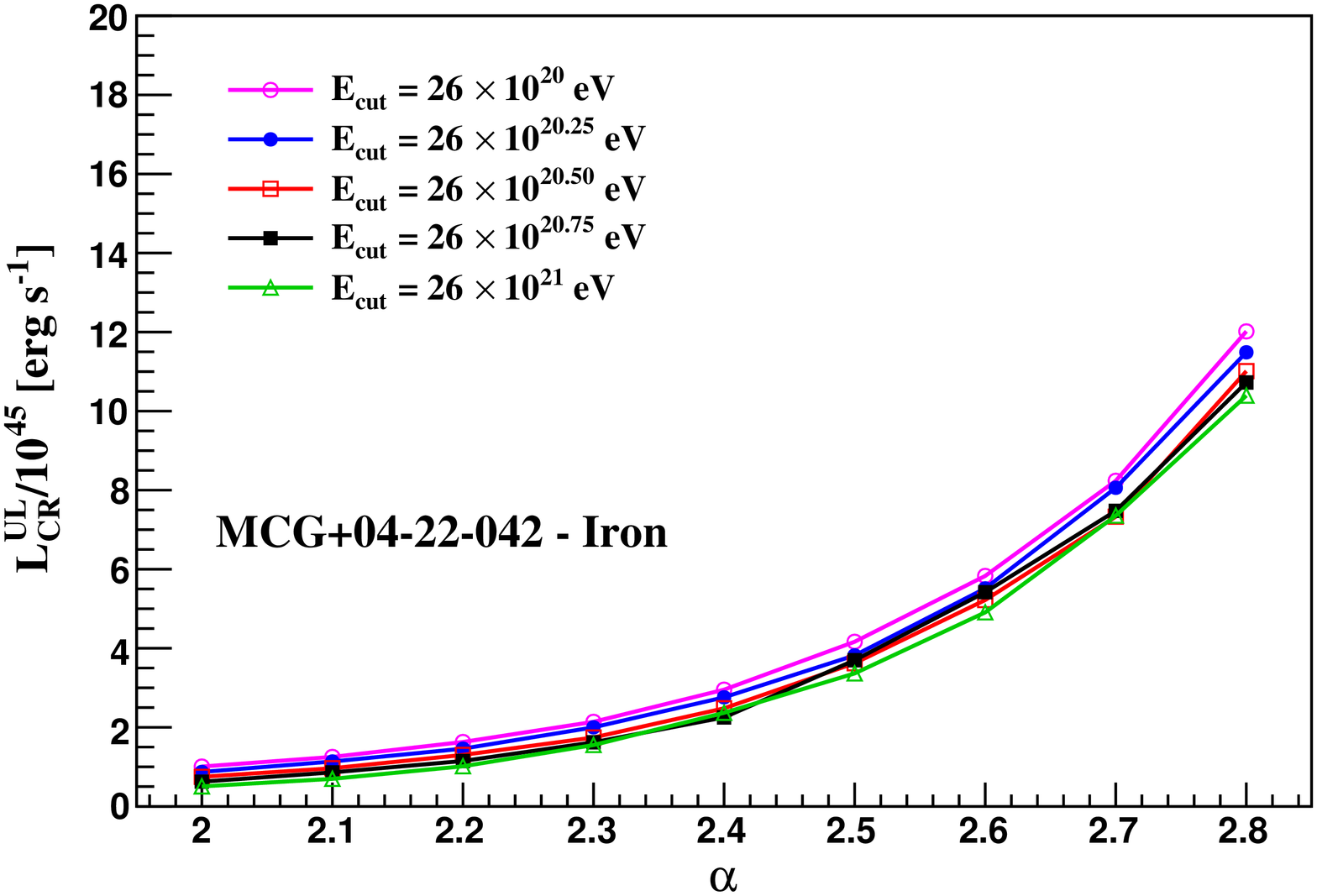}
  \includegraphics[angle=90,width=0.42\textwidth]{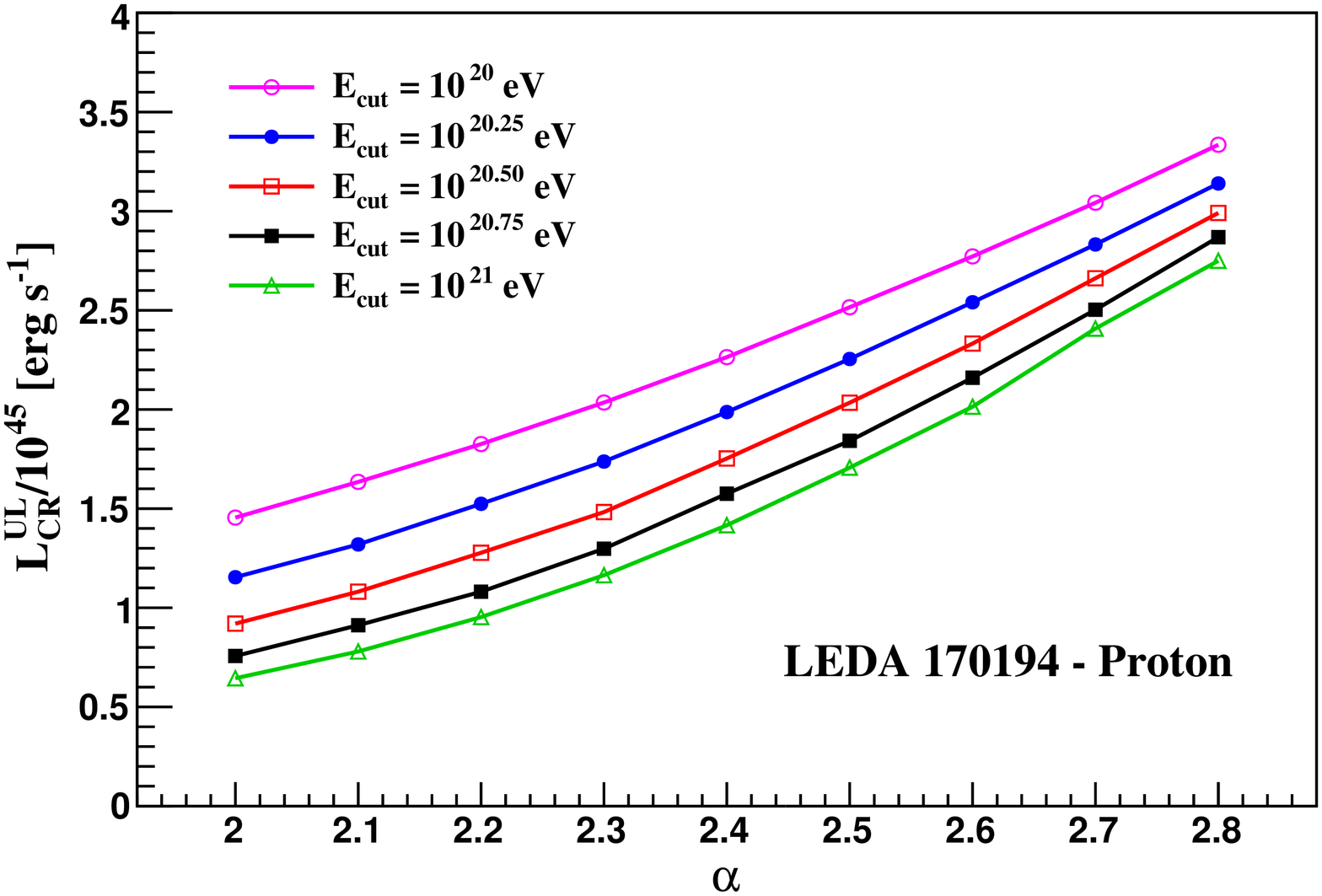}
  \includegraphics[angle=90,width=0.42\textwidth]{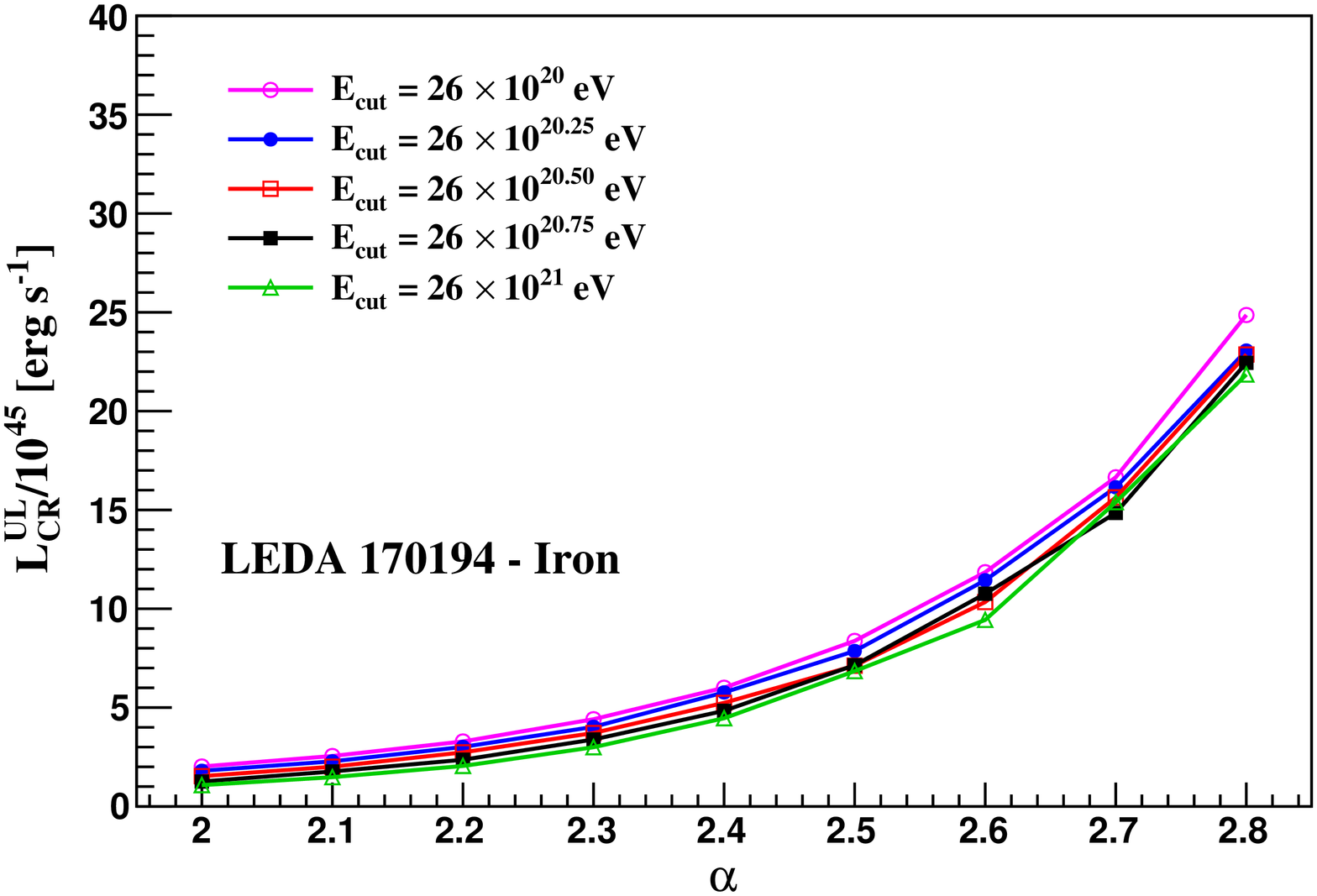}
  \includegraphics[angle=90,width=0.42\textwidth]{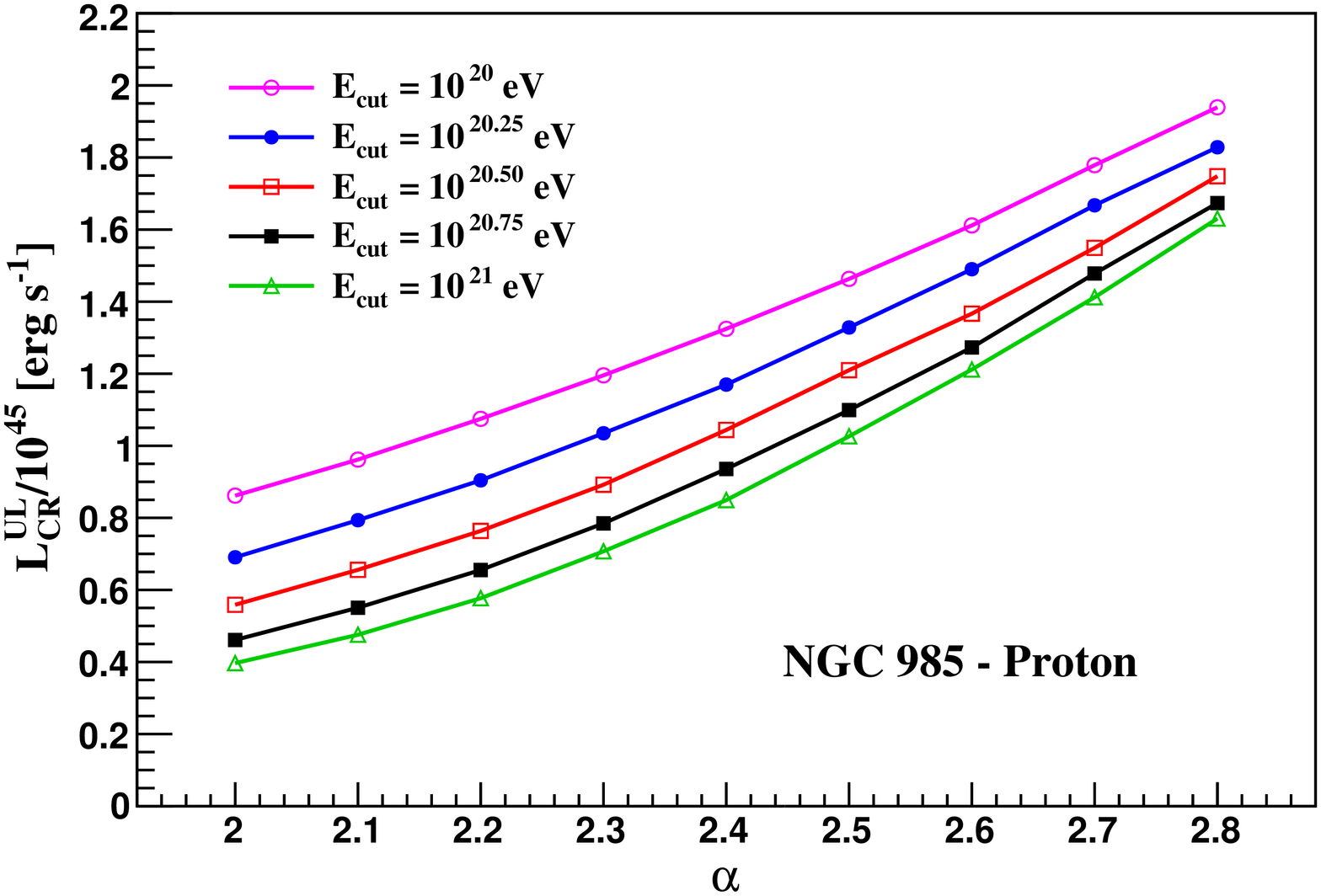}
  \includegraphics[angle=90,width=0.42\textwidth]{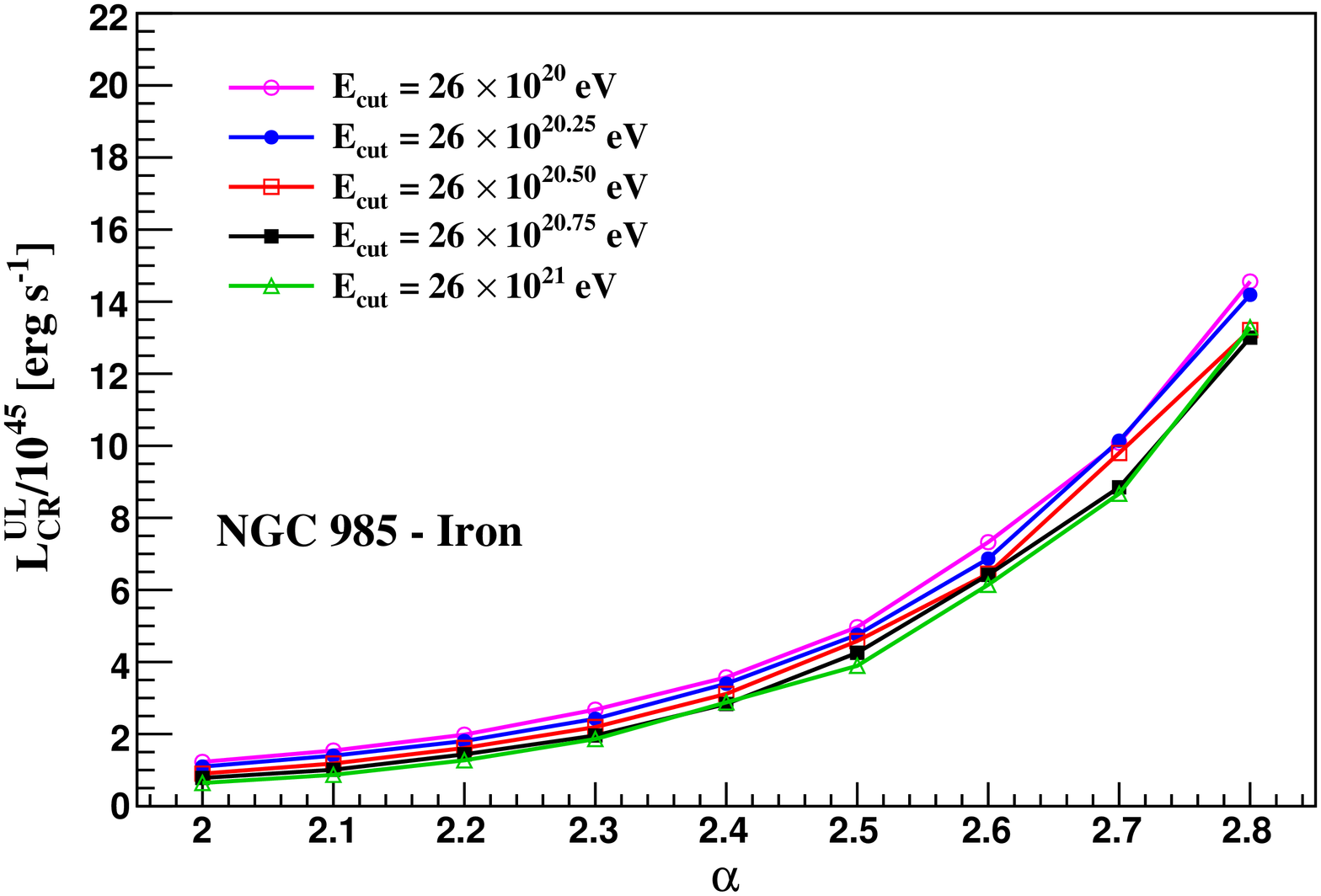}
  \includegraphics[angle=90,width=0.42\textwidth]{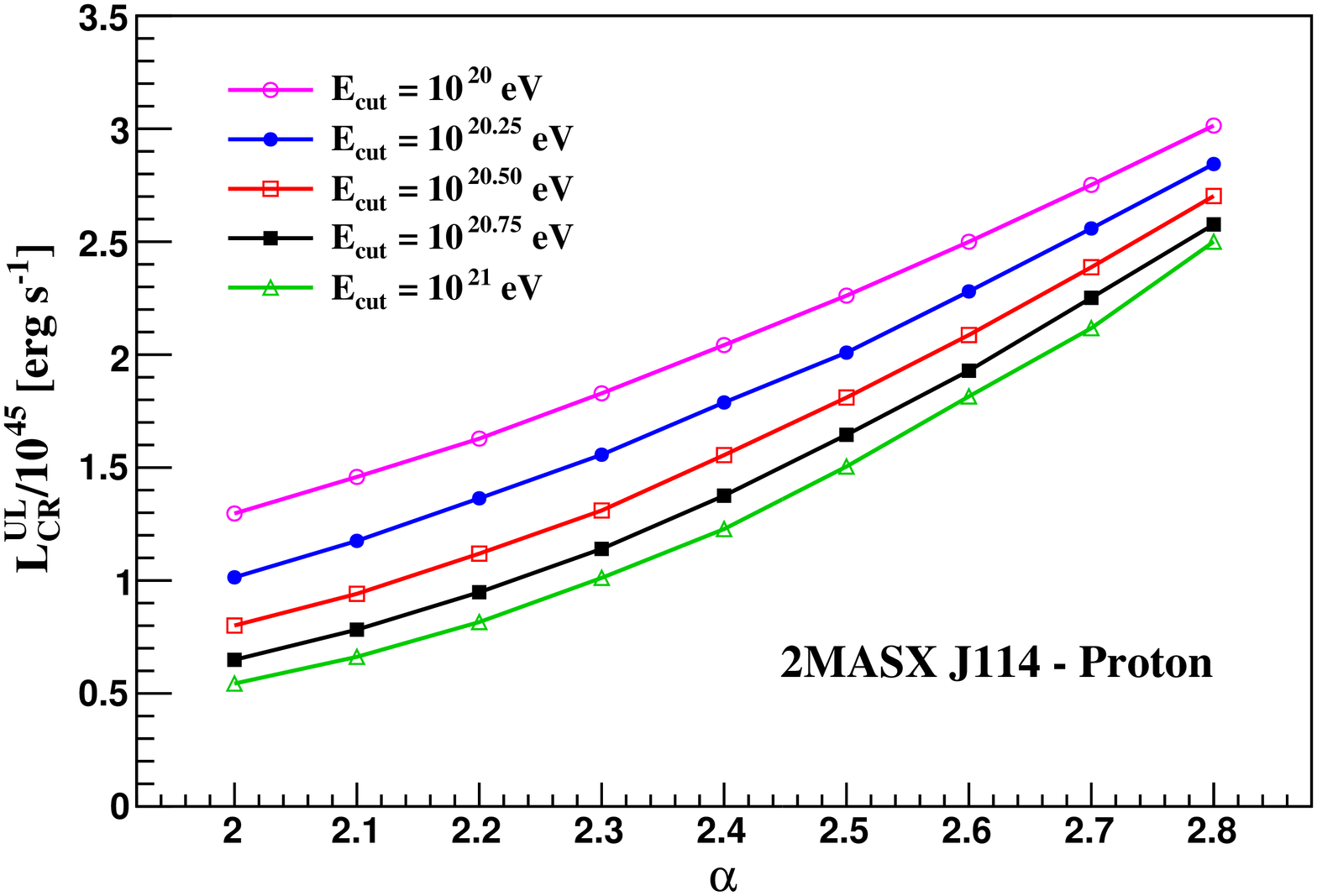}
  \includegraphics[angle=90,width=0.42\textwidth]{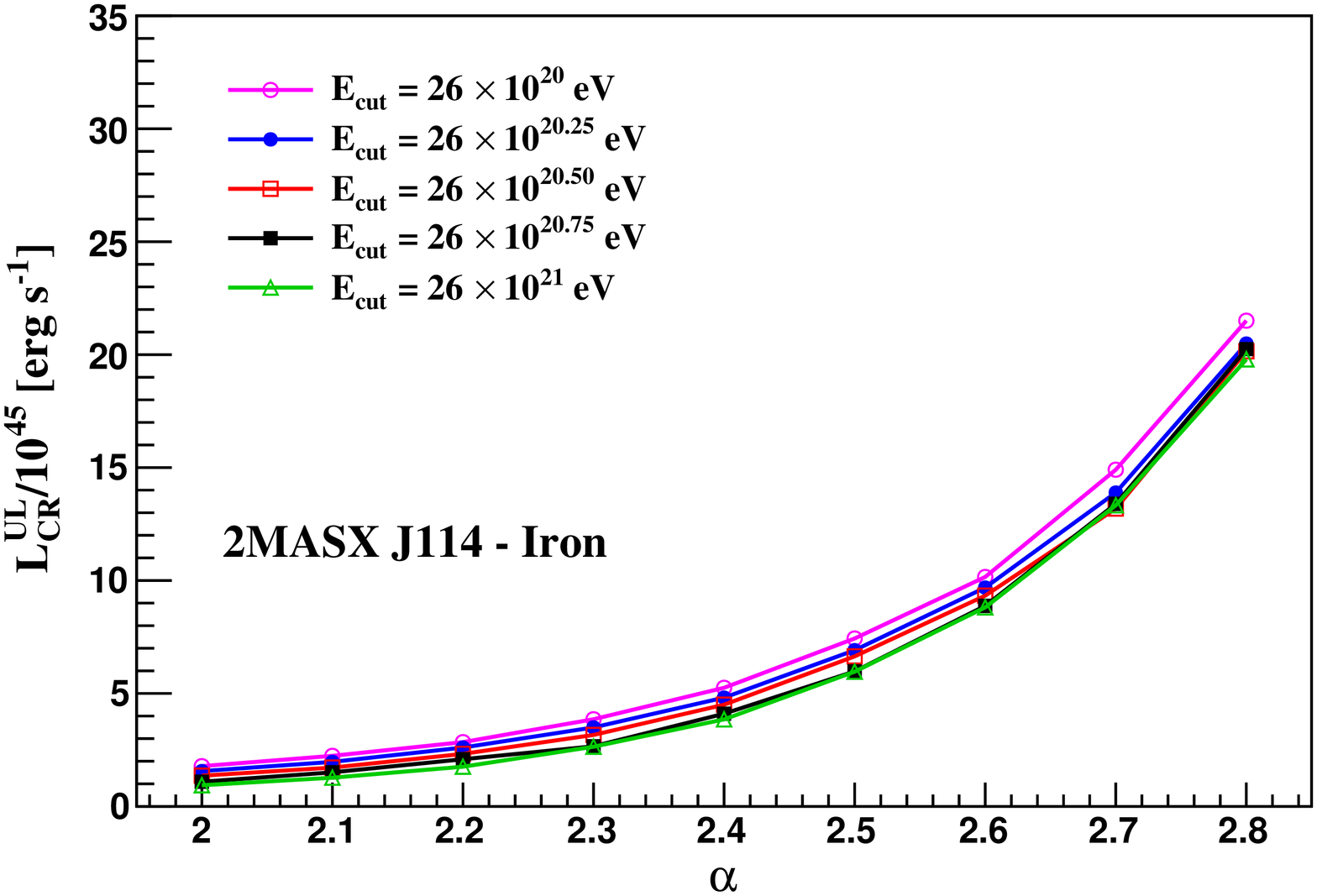}
  \includegraphics[angle=90,width=0.42\textwidth]{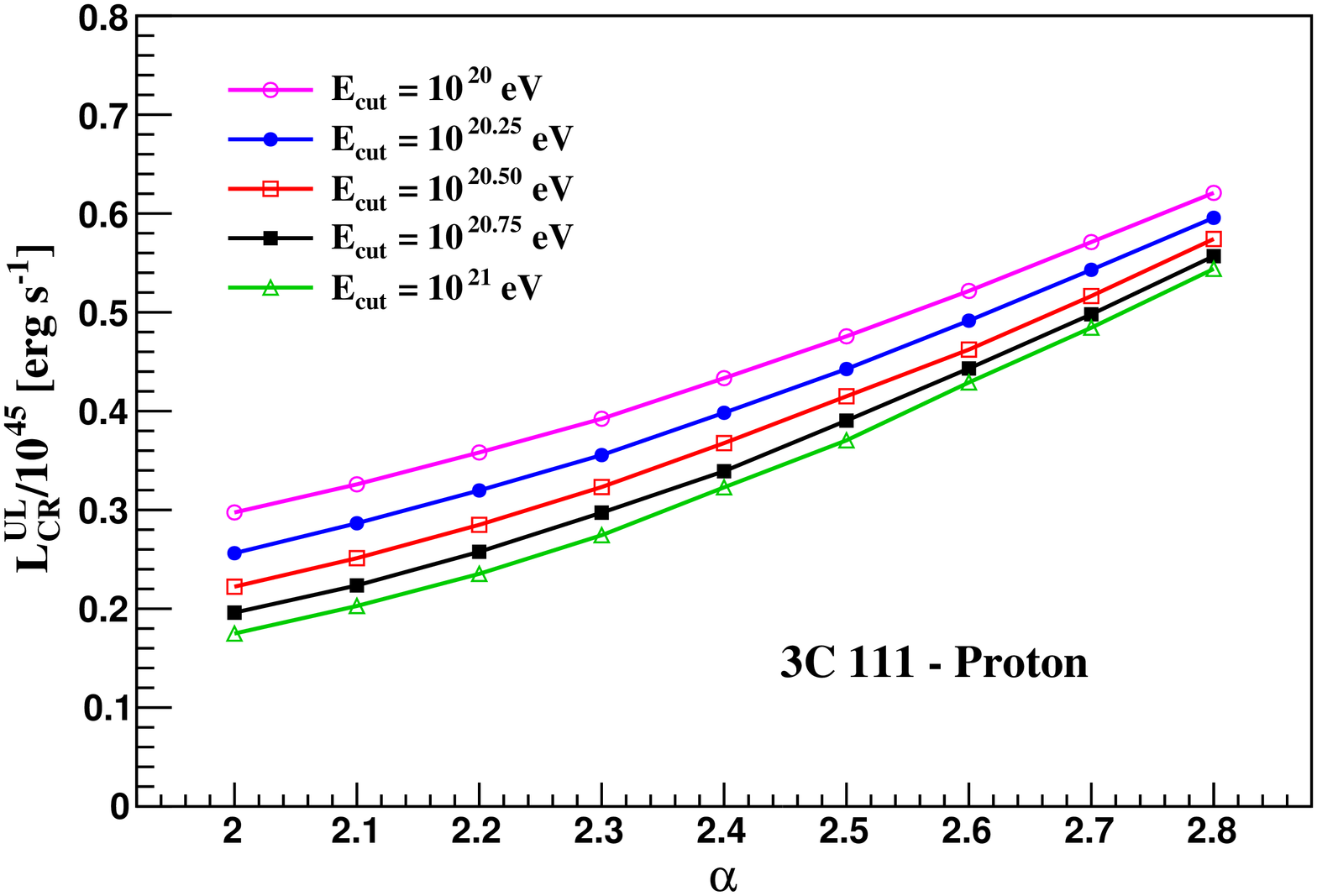}
  \includegraphics[angle=90,width=0.42\textwidth]{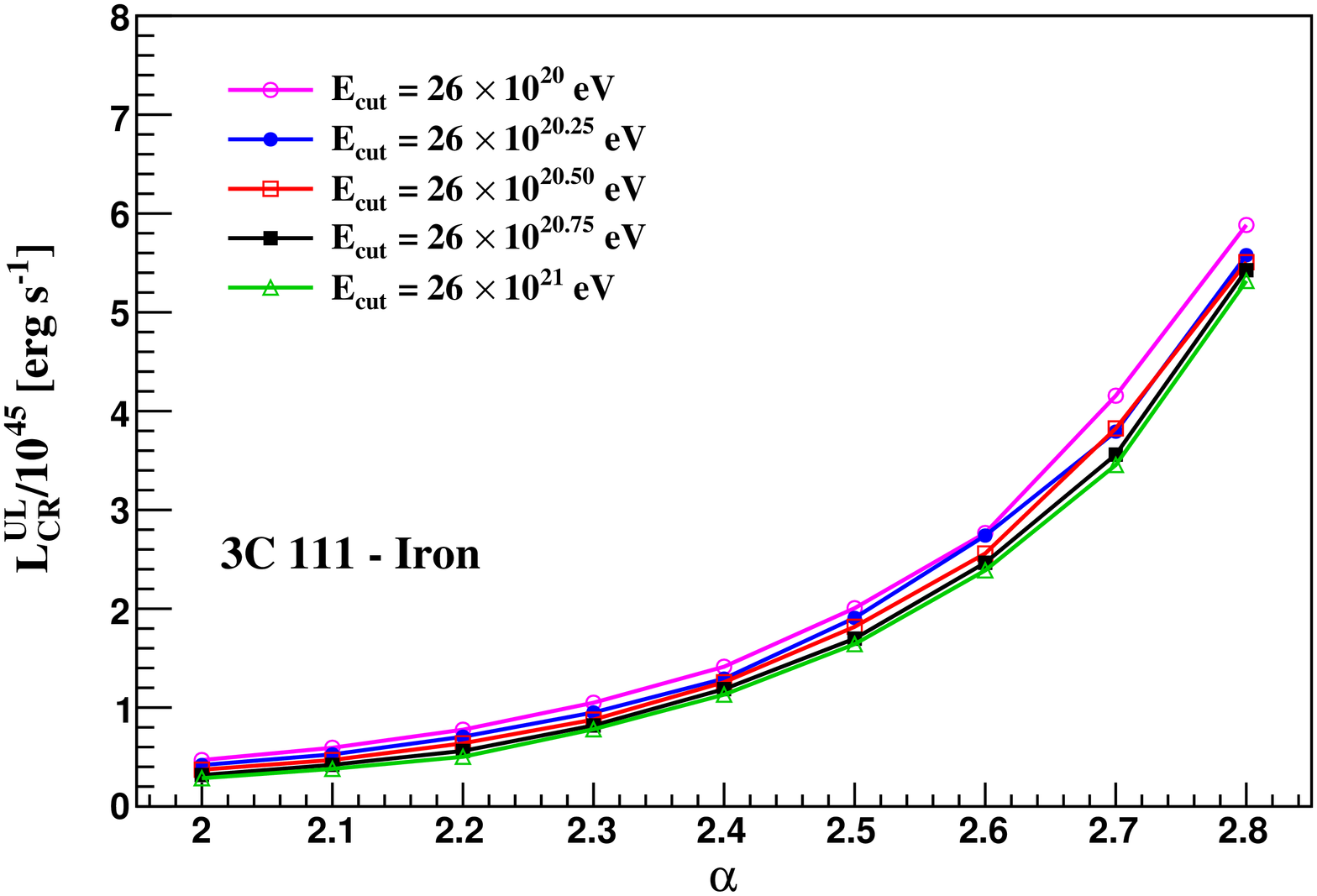}
  \caption{Upper limit on the total and proton cosmic-ray luminosity as inferred from the gamma-ray observations of the sources 3C 111, J11454045-1827149,
LEDA 170194, NGC 985, and MCG+04-22-042 as a function of the spectral index, for five values of the cutoff energy. Left: Primary proton. Right: Primary iron
nuclei.}
\label{fig:upper:total:luminosity}
\end{figure}

\begin{figure}
  \centering
  \includegraphics[angle=90,width=0.42\textwidth]{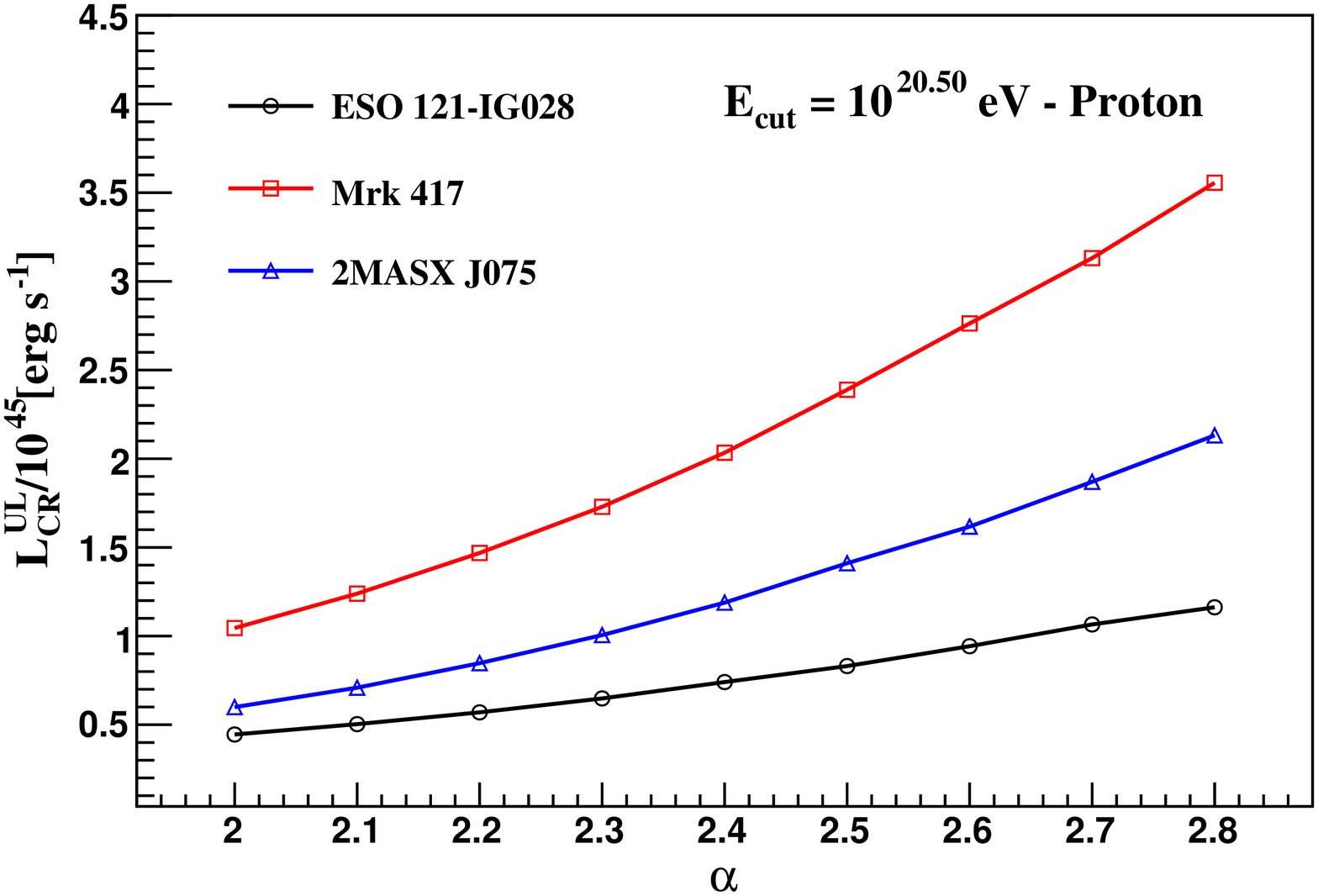}
  \includegraphics[angle=90,width=0.42\textwidth]{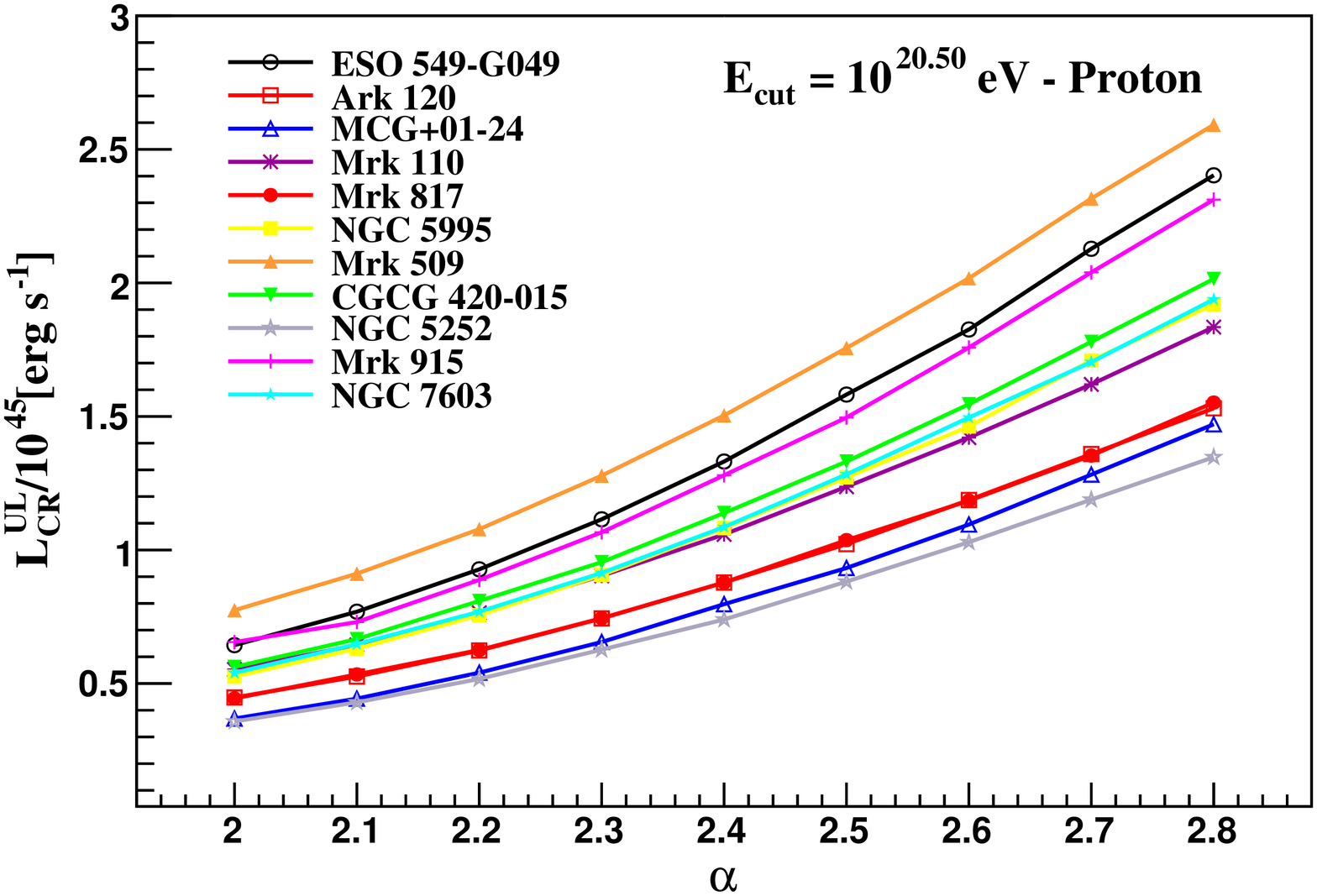}
  \includegraphics[angle=90,width=0.42\textwidth]{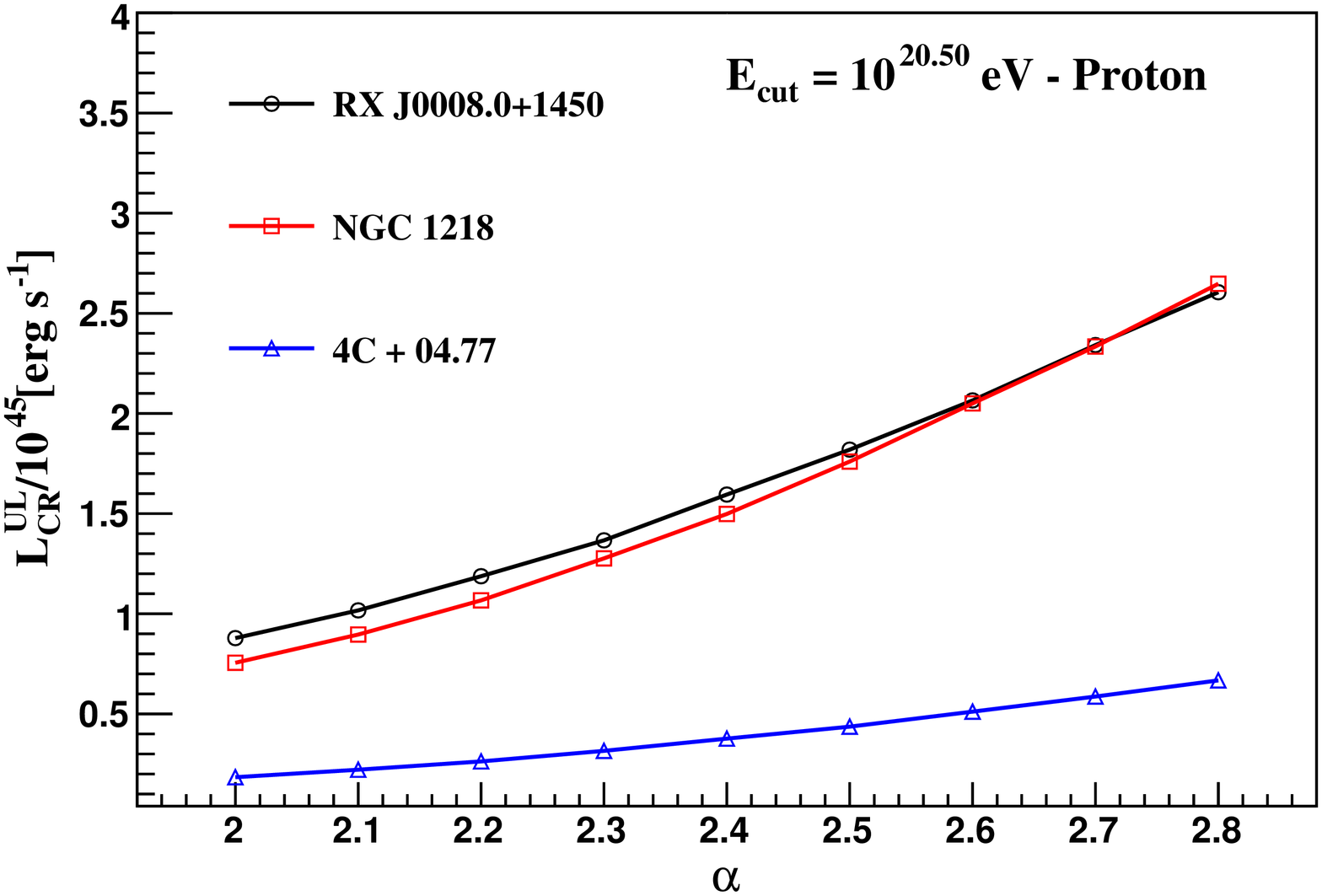}
  \includegraphics[angle=90,width=0.42\textwidth]{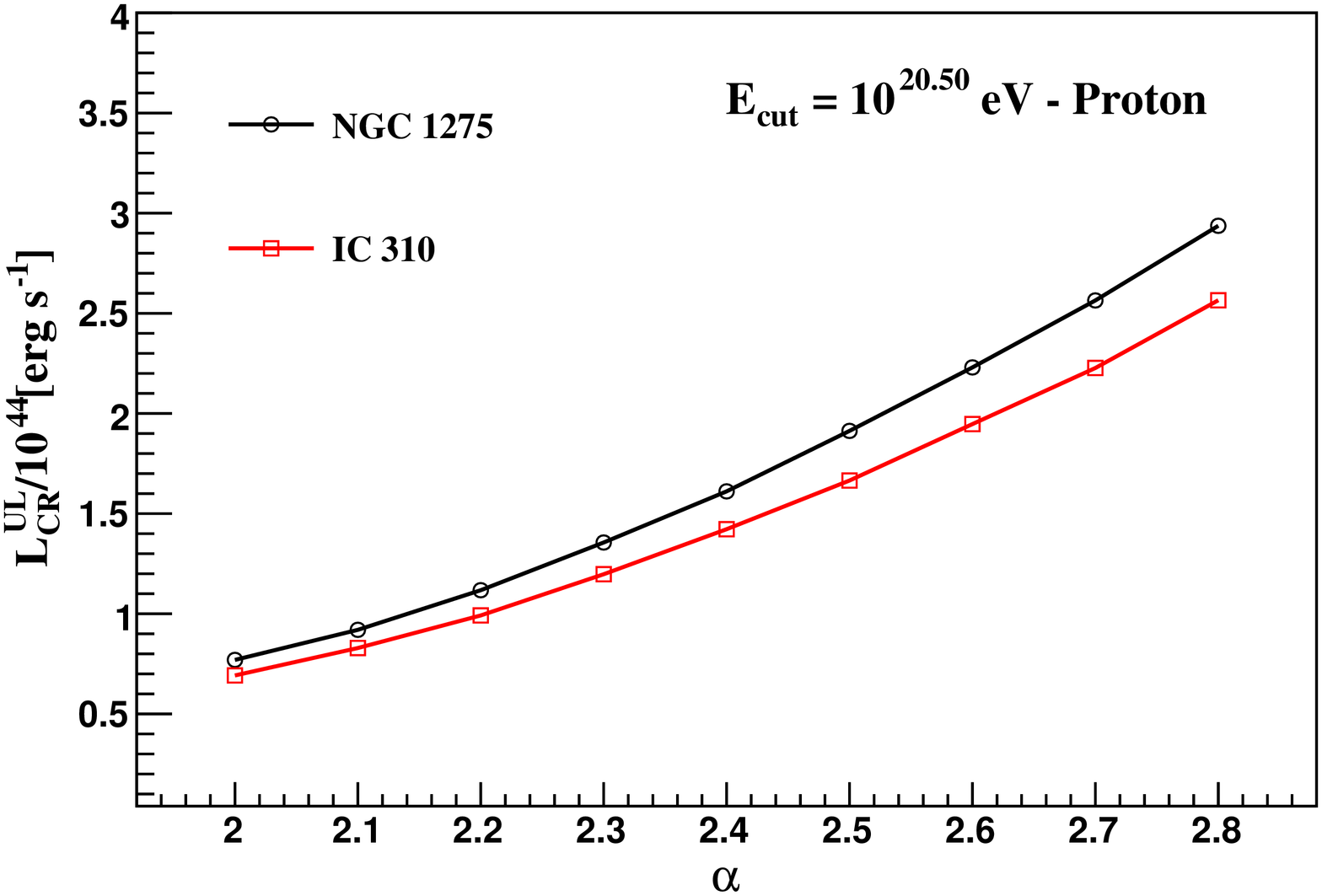}
  \caption{Upper limit on the proton luminosity for most of the sources shown in tables ~\ref{tab:1} - ~\ref{tab:3} as a function of the spectral index
and for $E_{cut} = 10^{20.50}$ eV.}
  \label{fig:upper:pr:luminosity}
\end{figure}

Figures~\ref{fig:upper:total:luminosity} and \ref{fig:upper:pr:luminosity} show the upper limit on the cosmic-ray luminosity as a
function of the spectral index for several $E_{cut}$ values and for the cases where protons and iron nuclei are considered as primaries. The
cases where iron nuclei are considered correspond to a conservative upper limit of the total UHECR luminosity for the reasons given above.
The results are summarized in~\cref{tab:1,tab:2,tab:3} for $\alpha = 2.3$ and $E_{cut} = Z \times 10^{20.5}$ eV. For some sources shown
in~\cref{tab:1,tab:2,tab:3}, the corresponding luminosity was not plotted because they show exactly the same behavior as the ones shown in
figures~\ref{fig:upper:total:luminosity} and \ref{fig:upper:pr:luminosity}. Nevertheless, the total or proton luminosities of all sources
are shown in~\cref{tab:1,tab:2,tab:3} for $\alpha = 2.3$ and $E_{cut} = Z \times 10^{20.5}$ eV.

\section{Conclusions}
\label{sec:conclusion}

In this paper twenty-eight AGNs were analyzed and an upper limit on the UHECR proton luminosity was set. For five out of twenty eight
sources, an upper limit on the UHECR total luminosity was set. Upper limit on the UHECR luminosity were calculated using the measured
upper limit on the GeV-TeV integral flux and a model on the propagation of UHECR particles from the source to Earth. It is important to emphasize
that the information of UHECR point sources is not directly accessed by UHECR experiments due to the lack of strong correlation between arrival
directions and source positions. Therefore the analysis discussed here represents an important step on the study of individual UHECR sources.

Moreover, the luminosity of UHECR is a fundamental restriction to the proposed models. According to reference~\cite{bib:restrictions}, several
restrictions can be applied to a model proposed to describe the generation of UHECR: a) geometry of the source , b) power of the source
(luminosity), c) radiation losses, d) interaction losses, e) emissivity and f) accompanying radiation. The analysis done here used the
accompanying radiation (GeV-TeV gamma-rays) to impose a limit on the power of the source.

The number of sources studied (28) for which this method offers a valid upper limit on the UHECR luminosity is still small. Nevertheless,
correlations of the UHECR upper limit on the luminosity with measured X-Ray and radio luminosities have been searched. No statistical
correlation has been found.

This paper presents the analysis of sources measured by the MAGIC and VERITAS telescopes and a selection of the extragalactic objects observed
by the Fermi-Lat experiment. The continuous operation of these experiments and the construction of the CTA Observatory will increase the number
of observed sources and enhance the sensitivity of the measurements significantly in the next years. At the same time, the sources for which an
upper limit on the UHECR luminosities can be set should be studied in details in other wavelengths. The combination of these multimessenger
information to come is certainly going to shed light on the puzzle of UHECR generation.

\appendix

\section{Upper limit on the total and proton cosmic-rays luminosity}
\label{sec:iron:max}

In case of the presence of different types of nuclei, the integral version of equation (\ref{eq:GammaFlux}) becomes,
\begin{equation}
I_{\gamma}(>E_{th}) = \frac{L_{CR}}{4 \pi D_s^2\ (1+z_s)} \ \sum_{A} f_A\ \frac{K_{\gamma}^A}{\langle E \rangle_0^A}\ %
\int_{E_{th}}^\infty dE_\gamma\ P_{\gamma}^A(E_\gamma),
\label{eq:GammaFluxAll}
\end{equation}
where the index $A$ corresponds to the different nuclear species injected by the source, $L_{CR}$ is the total cosmic-ray luminosity,
and $f_A=L_A/L_{CR}$ is the fraction of the total luminosity for nuclear type $A$. Note that by definition $f_A$ can take values
between 0 and 1 and also $\sum_A f_A = 1$.

In case that the gamma-ray observations provide a valid upper limit, an upper limit on the total cosmic-ray luminosity can be
obtained from equation (\ref{eq:GammaFluxAll}),
\begin{equation}
L_{CR}^{UL} = \frac{4\pi\ D^{2}_s\ (1+z_s)}{ {\mathop{\displaystyle \sum_A f_A\ \frac{K_{\gamma}^A}{\langle E \rangle_0^A}
\int_{E_{th}}^{\infty} dE_\gamma\ P_{\gamma}^A(E_{\gamma})}}}\ I_{\gamma}^{UL}(> E^{th}_{\gamma}),
\label{eq:CRULAll}
\end{equation}

Equation (\ref{eq:CRULAll}) can be rewritten as,
\begin{equation}
L_{CR}^{UL} = \frac{4\pi\ D^{2}_s\ (1+z_s)\ I_{\gamma}^{UL}(> E^{th}_{\gamma})}%
{{\mathop{\displaystyle f_{pr}\ \frac{K_{\gamma}^{pr}}{\langle E \rangle_0^{pr}}
\int_{E_{th}}^{\infty} dE_\gamma\ P_{\gamma}^{pr}(E_{\gamma})+\sum_{A \neq pr} f_A\ \frac{K_{\gamma}^A}{\langle E \rangle_0^A}
\int_{E_{th}}^{\infty} dE_\gamma\ P_{\gamma}^A(E_{\gamma})}}}.
\label{eq:CRULAllPr}
\end{equation}
The second term in the denominator of equation (\ref{eq:CRULAllPr}) is always positive and then if $L^{UL}_{pr} = f_{pr} L_{CR}^{UL}$
it is easy to see that,
\begin{equation}
L_{pr}^{UL} \leq \frac{4\pi\ D^{2}_s\ (1+z_s)}%
{{\mathop{\displaystyle \frac{K_{\gamma}^{pr}}{\langle E \rangle_0^{pr}}
\int_{E_{th}}^{\infty} dE_\gamma\ P_{\gamma}^{pr}(E_{\gamma})}}}\ I_{\gamma}^{UL}(> E^{th}_{\gamma}),
%
%
\end{equation}
which shows that for the case of a pure proton composition injected by the source, equation (\ref{eq:CRUL}) gives a conservative
upper limit on the proton luminosity of the source.

Let us define the following parameter,
\begin{equation}
M_A=\frac{K_{\gamma}^{A}}{\langle E \rangle_0^{A}} \int_{E_{th}}^{\infty} dE_\gamma\ P_{\gamma}^{A}(E_{\gamma}).
\label{eq:M}
\end{equation}
As discussed previously, the GeV-TeV gamma-ray flux is inversely proportional to the mass of the injected primary particles. If
$\langle E \rangle_0^A$ is an increasing function of $A$, the parameter $M_A$ is a decreasing function of $A$. Note that,
in general, this condition is fulfilled for the physically motivated scenarios in which the maximum energy of the different primaries
increases with charge number. Then, the denominator of equation (\ref{eq:CRULAll}) can be written as,
\begin{equation}
\xi=\sum_{A} f_A\ M_A.
\label{eq:xi}
\end{equation}
Therefore, $\xi$ can be rewritten in the following form,
\begin{equation}
\xi=f_{fe}\ M_{fe} + \sum_{A\neq fe} f_A\ M_A = M_{fe} + \sum_{A\neq fe} f_A\ (M_A-M_{fe}),
\label{eq:xif}
\end{equation}
where it is used that $f_{fe}=1-\sum_{A\neq fe} f_A$. If iron nuclei are the heaviest element accelerated by the sources, then
$M_A-M_{fe} \geq 0$. From equation (\ref{eq:xif}) it can be seen that the minimum value of $\xi$ is attained when $f_A=0$ for all
$A \neq fe$ and $f_{fe}=1$. Therefore,
\begin{equation}
L_{CR}^{UL} \leq \frac{4\pi\ D^{2}_s\ (1+z_s)}%
{{\mathop{\displaystyle \frac{K_{\gamma}^{fe}}{\langle E \rangle_0^{fe}}
\int_{E_{th}}^{\infty} dE_\gamma\ P_{\gamma}^{fe}(E_{\gamma})}}}\ I_{\gamma}^{UL}(> E^{th}_{\gamma}),
%
%
\end{equation}
i.e.~a conservative upper limit on the total cosmic-ray luminosity is obtained from equation (\ref{eq:CRUL}) considering a source
that injects a pure composition of iron nuclei.

\acknowledgments

RCA thanks CAPES. VdS thanks the support of the Brazilian population via CNPq and FAPESP (2012/22540-4). ADS is member of the Carrera del Investigador Cient\'{\i}fico of CONICET, Argentina. The work of ADS is supported by CONICET PIP 114-201101-00360 and ANPCyT PICT-2011-2223, Argentina. The authors thank the Pierre Auger Collaboration for permission to use their data prior to journal publication.

\begin{landscape}

\begin{table}[p]
\centering
\begin{tabular}{|c|c|c|c|c|c|}
\hline \textbf{Source name} & \textbf{D} [Mpc] &  \textbf{UL:} $\mathcal{F}$ ($>0.1$ GeV) &  $\mathbf{L_{pr}^{UL}}$ (Proton) & $\mathbf{L_{CR}^{UL}}$ (Total) \\
                &             &        [$10^{-9}$ ph cm$^{-2}$ s$^{-1}$]   &   [erg s$^{-1} \times 10^{45}$]&   [erg s$^{-1} \times 10^{45}$]           \\ \hline
  Mrk 1018      &  181.5      &  2.1     &      1.04      &             -                    \\ \hline
  NGC 985       &   184.7     &  1.8     &      1.03       &              2.19                 \\ \hline
  NGC 1142      &   121.5     &  1.1     &      0.49       &              -                  \\ \hline
  2MASX J07595347+2323241  & 127.7  &  2.2     &      1.01       &              -               \\ \hline
  Mrk 704       &  130        &  2.0     &      0.91      &              -                       \\ \hline
  MCG+04-22-042 &  143.6      &  1.6     &       0.75     &              1.74                      \\ \hline
  Mrk 417       &  147.4      &  3.7     &       1.72      &              -                   \\ \hline
  ESO 121-IG028 &  177.8      &  1.2     &       0.64     &              -                    \\ \hline
  ESO 549-G049  &  111.1      &  2.5     &        1.11    &              -                     \\ \hline
  CGCG 420-015  &   124.8     &  2.1     &        0.95    &              -                       \\ \hline
  Ark 120       &   139.7     &  1.6     &        0.74    &              -                    \\ \hline
  MCG-01-24-012 &  89.0       &  1.5     &        0.65    &              -                        \\ \hline
  Mrk 110       &    156      &  1.9     &        0.90    &              -                        \\ \hline
  2MASX J11454045-1827149 & 150.7   &  2.8     &        1.30     &        3.16                        \\ \hline
  LEDA 170194   &  167.7      &  3.1     &        1.48     &              3.71                     \\ \hline
  NGC 5252      &   108.4     &  1.4     &        0.62    &              -                   \\ \hline
  Mrk 817       &   141.5     &  1.6     &        1.72     &              -                      \\ \hline
  NGC 5995      &  118.1      &  2.0     &        0.90    &              -                          \\ \hline
  Mrk 509       &  151.6      &  2.7     &        1.27     &              -                        \\ \hline
  Mrk 520       &  115.5      &  2.2     &        0.98    &              -                          \\ \hline
  Mrk 915       &    104      &  2.4     &        1.06     &              -                        \\ \hline
  NGC 7603      &  126.5      &  2.0     &        0.91    &              -                           \\ \hline
\end{tabular}
\caption{Column 1 shows the source name. Column 2 and 3 were taken from references~\cite{bib:fermi:cat:1,bib:fermi:cat:sey}. They show the distance of the
source from Earth and the 95\% CL upper limits on the gamma-ray flux above 0.1 GeV as measured by the FERMI-LAT Observatory. Column 4 shows the upper
limit on the proton luminosity for each source as calculated in this paper. Column 5 shows the upper limit on the total UHECR luminosity for each
source as calculated in this paper.}
\label{tab:1}
\end{table}

\begin{table}[p]
\centering
\begin{tabular}{|c|c|c|c|c|c|c|c|}
\hline \textbf{Source name} & \textbf{D} [Mpc]  & \textbf{UL:} $\mathcal{F}$ ($ > E_{th}$) & $\mathbf{E_{th}}$  & \textbf{Measured by}& $\mathbf{L_{pr}^{UL}}$ (Proton) & $\mathbf{L_{CR}^{UL}}$ (Total) \\
           &        &   [ph cm$^{-2}$ $s^{-1}$]    &      [GeV]   &      &  [erg s$^{-1} \times 10^{45}$]   & [erg s$^{-1} \times 10^{45}$]      \\\hline
  3C 111   & 196        & 2.5 $\times$ $10^{-12}$ &   300    & VERITAS         & 0.32   &    0.87        \\ \hline
  NGC 1275 & 75         & 5.11 $\times$ $10^{-12}$ &   190     & VERITAS       & 0.13   &     -       \\ \hline
  IC 310   & 78.5375    & 3.1 $\times$ $10^{-12}$ &   300    & MAGIC           & 0.11   &     -          \\ \hline
 \end{tabular}
\caption{Column 1 shows the source name. Column 2 and 3 were taken from reference~\cite{bib:magic,bib:veritas}. They show the distance of the source from
Earth and the 99\% CL VERITAS (95\% C.L. MAGIC) upper limits on the gamma-ray flux above $E_{th}$. Column 4 shows the energy threshold ($E_{th}$) of the
measurement. Column 5 shows the gamma-ray observatory that observed the source. Column 6 shows the upper limit on the proton luminosity for each
source as calculated in this paper. Column 7 shows the upper limits on the total UHECR luminosity for each source as calculated in this paper.}
\label{tab:2}
\end{table}

\begin{table}[p]
\centering
\begin{tabular}{|c|c|c|c|}
\hline \textbf{Sources}& \textbf{D} [Mpc] & \textbf{UL:} $\mathcal{F}$ ($10 < E < 100$ GeV) & $\mathbf{L_{CR}^{UL}}$ (Proton)   \\
                     &                   &  [ph cm$^{-2}$ s$^{-1}$]     &  [erg s$^{-1} \times 10^{45}$]               \\ \hline
    NGC 1218         &       120.83      &  1.47 $\times$ $10^{-10}$   &      1.27                                        \\ \hline
    4C+04.77         &       112.5       &  8.59 $\times$ $10^{-11}$   &      0.31                                       \\ \hline
    RX J0008.0+1450  &       187.5       &  1.22 $\times$ $10^{-10}$   &        1.36                                      \\ \hline
\end{tabular}
\caption{Column 1 shows the source name. Column 2 and 3 were taken from reference~\cite{bib:fermi:cat:1,bib:fermi:cat:sey}. They show the distance
of the source from Earth and the  95\% CL upper limits on the gamma-ray flux with energy $10 < E_\gamma < 100$ GeV as measured by the FERMI-LAT
Observatory. Column 4 shows the upper limit on the proton luminosity for each source as calculated in this paper. No upper limit on the total UHECR
luminosity is shown because the gamma-ray upper limits were not restrictive enough for these sources.}
\label{tab:3}
\end{table}

\end{landscape}

\end{document}